\documentclass{article}

\usepackage{arxiv}

\usepackage[utf8]{inputenc} 
\usepackage[T1]{fontenc}    
\usepackage{lmodern}        
\usepackage{hyperref}       
\usepackage{url}            
\usepackage{booktabs}       
\usepackage{amsfonts}       
\usepackage{nicefrac}       
\usepackage{microtype}      
\usepackage{lipsum}
\usepackage{graphicx}

\title{The Impact of the U.S. Census Disclosure Avoidance System on Redistricting and Voting Rights Analysis}

\author{
    Christopher T. Kenny
   \\
    Department of Government \\
    Harvard University \\
  Cambridge, MA \\
  \texttt{\href{mailto:christopherkenny@fas.harvard.edu}{\nolinkurl{christopherkenny@fas.harvard.edu}}} \\
   \And
    Shiro Kuriwaki
   \\
    Department of Government \\
    Harvard University \\
  Cambridge, MA \\
  \texttt{\href{mailto:kuriwaki@g.harvard.edu}{\nolinkurl{kuriwaki@g.harvard.edu}}} \\
   \And
    Cory McCartan
   \\
    Department of Statistics \\
    Harvard University \\
  Cambridge, MA \\
  \texttt{\href{mailto:cmccartan@fas.harvard.edu}{\nolinkurl{cmccartan@fas.harvard.edu}}} \\
   \And
    Evan Rosenman
   \\
    Harvard Data Science Initiative \\
    Harvard University \\
  Cambridge, MA \\
  \texttt{\href{mailto:erosenm@fas.harvard.edu}{\nolinkurl{erosenm@fas.harvard.edu}}} \\
   \And
    Tyler Simko
   \\
    Department of Government \\
    Harvard University \\
  Cambridge, MA \\
  \texttt{\href{mailto:tsimko@g.harvard.edu}{\nolinkurl{tsimko@g.harvard.edu}}} \\
   \And
    Kosuke Imai
    \thanks{We thank many people who provided us with comments, criticisms, and
suggestions on the initial draft of this paper, which was prepared in
response to the Census Bureau's April 2021 request for public feedback. We
also thank Bruce
Willsie of L2, Inc.~for providing voterfile data.
Code and datasets for replication are available at \url{https://doi.org/10.7910/DVN/TNNSXG}.}
   \\
    Department of Government and Department of Statistics \\
    Harvard University \\
  Cambridge, MA \\
  \texttt{\href{mailto:imai@harvard.edu}{\nolinkurl{imai@harvard.edu}}} \\
  }

\usepackage{placeins}
\usepackage{booktabs}
\usepackage{longtable}
\usepackage{array}
\usepackage{multirow}
\usepackage{wrapfig}
\usepackage{float}
\usepackage{colortbl}
\usepackage{pdflscape}
\usepackage{tabu}
\usepackage{threeparttable}
\usepackage{threeparttablex}
\usepackage[normalem]{ulem}
\usepackage{makecell}
\usepackage{xcolor}

\begin{document}
\maketitle

\def\tightlist{}

\begin{abstract}
The U.S. Census Bureau plans to protect the privacy of 2020 Census respondents
through its Disclosure Avoidance System (DAS), which attempts to achieve
differential privacy guarantees by adding noise to the Census microdata.
We empirically investigate the impact of DAS on redistricting and voting
rights analysis under a likely scenario in which practitioners---map drawers
and analysts alike---treat the DAS-protected data ``as-is'' without accounting
for the DAS mechanism. We apply redistricting simulation and analysis methods
to the DAS-protected 2010 Census data and compare the results with those
obtained based on the original Census data. Our analysis of federal, state,
and local redistricting cases shows that the most recently released levels of
noise would have profound effects on
redistricting in the United States. First, we find that DAS has a tendency
to transfer population across geographies in ways that artificially reduce
racial and partisan heterogeneity. In some cases,
DAS-protected data can yield the number of majority-minority
districts that is different from the one obtained under the original Census data.
These differences can lead to unpredictable changes in the
analysis of partisan and racial gerrymanders. Second, the injected noise makes it
impossible to accurately comply with the \emph{One Person, One Vote} principle as
currently interpreted and implemented. Finally, we find that the DAS does not
degrade the overall predictive accuracy of an individual's race based on the voter
files and Census data. Yet, the individual-level race predictions based on the
DAS data differ significantly from those based on the original Census data, potentially
affecting the analysis of redistricting cases especially for local offices. We conclude
with implications of our findings for future redistricting and voting rights
analysis under the privacy-protected Census data.
\end{abstract}

\keywords{
    Census
   \and
    Redistricting
   \and
    BISG
   \and
    Differential privacy
   \and
    TopDown algorithm
   \and
    One Person One Vote
  }

\clearpage
\begin{center}
\begin{minipage}{0.8\linewidth}
\tableofcontents
\end{minipage}
\end{center}

\vspace{0.5in}

\newgeometry{margin=1.5in}

\hypertarget{introduction}{%
\section{Introduction}\label{introduction}}

In preparation for the official release of the 2020 Census data, the United States Census Bureau has developed its Disclosure Avoidance System (DAS) to prevent Census responses from being linked to specific individuals \cite{abow:etal:20}.
The DAS is based on differential privacy technology, which adds a certain amount of random noise to the raw Census counts.
The Bureau has been required by law to prevent the disclosure of information about Census participants (13 U.S.C. \(\mathsection\) 9) and has implemented disclosure avoidance methods since 1960.
Their decision to use differential privacy starting in 2020, however, has been controversial.
Some scholars have voiced concerns about the potential negative impacts of noisy data on public policy and social science research, which critically rely upon the Census data \cite{rugg:etal:19, nas:20, ripe:kugl:rugg:20, sant:etal:20, muel:sant:21}.

In this paper, we empirically evaluate the impact of the DAS on redistricting and voting rights analysis.
Once released as part of the 2020 Census data later this year, states are likely to use the DAS-protected P.L.
94-171 redistricting data to redraw their district boundaries of Congressional and other federal and local electoral offices.
It is therefore of paramount importance to examine how the DAS affects redistricting analysis and process.
To this end, on April 28, 2021, the Census Bureau released Demonstration Data that apply the new DAS to the results of the 2010 decennial Census.
We conduct our empirical evaluation under a likely scenario, in which practitioners---map drawers and analysts alike---treat this DAS-protected data ``as-is,'' as they have done in the past, without accounting for the DAS noise generation mechanism.
Our evaluation is based on the final iteration of the Demonstration Data before the Census plans to release the 2020 P.L.
94-171 data in August 2021.

We employ a set of recently-developed simulation methods that can generate large numbers of realistic redistricting maps under a set of legal and other relevant constraints, including contiguity, compactness, population parity, and preservation of communities of interest and counties \cite{chen2013, cart:etal:19, deford2019, autry2020, fifi:etal:20, mcca:imai:20, kenn:etal:21}.
These simulation methods have been extensively used by expert witnesses in recent court cases on redistricting, including \emph{Common Cause v. Lewis} (2020), \emph{Rucho v. Common Cause} (2019), \emph{Ohio A. Philip Randolph Institute v. Householder} (2020), and \emph{League of Women Voters of Michigan v. Benson} (2019).

By analyzing data from several states across federal, state, and local settings, we find that the DAS would have profound effects on redistricting in the United States.
First, we find significant biases in the DAS-protected data along racial and partisan lines.
The DAS has a tendency to transfer population across geographies in ways that artificially reduce racial and partisan heterogeneity.
In some cases, these perturbations can even lead to a change in the number of majority-minority districts in a given area, using a currently-standard measurement (e.g.~\emph{Thornburg v. Gingles} 1986, \emph{Shaw v. Reno} 1993, \emph{Bartlett v. Strickland} 2009, \emph{Shelby County v. Holder} 2013).

Second, we find that the noise introduced by the DAS can prevent map drawers from creating districts of equal population according to current statutory and judicial standards.
For example, over the past half century the Supreme Court has firmly established the principle of \emph{One Person, One Vote}, requiring states to minimize the population difference across districts based on the Census data (e.g.~\emph{Karcher v. Daggett} 1983).
In many cases, actual deviations from equal population, as measured using the original Census data, will be several times larger than as reported under the DAS-protected data.
The magnitude of this problem increases for smaller districts, such as state legislative districts and school boards.

Third, we find that the noise-induced DAS data do not degrade the overall prediction accuracy of each individual's race based on the Bayesian Improved Surname Geocoding (BISG) methodology, which combines the Census block-level proportion of each race with a voter's name and address \cite{fisc:frem:06,elli:etal:09,imai:khan:16}.
Redistricting analysis for voting rights cases often necessitates such individualized prediction, because most states' voter lists do not include individual race.
However, we also show that the DAS-protected data can substantially alter individual-level race predictions constructed from voter names and addresses.
These changes can have a significant impact the analysis of \emph{local} redistricting cases.
In a re-analysis of a recent Voting Rights Act case, \emph{NAACP v. East Ramapo School District.}, we find that predictions generated using DAS-protected data under-predict minority voters and result in fewer majority-minority districts.

Finally, we will discuss the implications of our findings for future redistricting and voting rights analysis under the privacy-protected Census data.
Based in part on an earlier version of this paper, along with input from others, the Census Bureau has decided to alter its DAS algorithm and address some of the aforementioned problems.
The Bureau now plans to further increase the privacy loss budget and modify the post-processing algorithm.\footnote{See the Census Bureau's June 9, 2021 press release: \url{https://www.census.gov/newsroom/press-releases/2021/2020-census-key-parameters.html}}
The Census Bureau claims that, on a national scale, this has lowered the total error at areas above the optimized block group, but has increased the amount of error introduced by the DAS at the block level.\footnote{See the Census Bureau's July 1, 2021 newsletter: \url{https://content.govdelivery.com/accounts/USCENSUS/bulletins/2e5e8a6}} New demonstration data based on this updated DAS algorithm will not be made available for at least several weeks after the 2020 Census data release.
Unfortunately, this makes it impossible to evaluate the impact of the finalized DAS algorithm at this time.
Nonetheless, our analysis identifies systematic biases in the last version of demonstration data that are publicly available before many states begin redistricting.
If the same flaws are not resolved in the 2020 Census data, they may have important ramifications for the upcoming redistricting cycle and for years to come.

Our paper represents the broadest look at the impact of the new DAS methodologies on redistricting use to date.
Prior research applied related redistricting simulation methodologies to simulated DAS data, as we do, but used an old version of the DAS algorithm rather than the latest demonstration data release \cite{MGGG:21}.
The authors make a valuable contribution in demonstrating the continued usability of weighted regressions for voting rights analysis.
While their primary focus is on the analysis of one county in one state, we cover several levels of redistricting across many states, which allows us to examine the consequences of DAS-induced error across a variety of contexts and use cases.

\hypertarget{overview-of-simulation-analysis}{%
\section{Overview of Simulation Analysis}\label{overview-of-simulation-analysis}}

As mentioned above, in April 2021, the Census Bureau has released Privacy-Protected Microdata Files based on the application of the DAS to the 2010 Census redistricting data.
The amount of noise in each file is controlled by the \emph{privacy loss budget,} denoted by \(\epsilon\).
The DAS-12.2 data are based on a relatively higher level of privacy loss budget (\(\epsilon = 12.2\)) to achieve the accuracy targets at the expense of greater privacy loss, whereas the DAS-4.5 data use a lower privacy loss budget at the expense of worse accuracy (\(\epsilon = 4.5\)).

In addition, the Census Bureau post-processes the noisy data in order to ensure that the resulting public release data are self-consistent (e.g., no negative counts) and certain aggregate statistics such as state-level total population counts are accurate.
These releases, alongside the originally released 2010 Census, allow us to analyze how DAS may affect redistricting.

For the purposes of evaluating the impact of the new DAS on redistricting plan-drawing and analysis, we generated nine sets of redistricting datasets for simulation, described in Table \ref{tab:simulations}.
We create precinct-level datasets that have three versions of total population counts: the original 2010 Census, the DAS-12.2 data, and the DAS-4.5 data.\footnote{We assemble 2010 Census data and obtain matches between blocks and districts using the R package \texttt{geomander} \cite{geomander}.
  We create corresponding block-level data by aggregating the privacy protected microdata using an R package developed for this project, \texttt{ppmf} \cite{ppmf}.}
These cases cover federal, state, and local offices across a diverse set of states.

In our modal analysis, we simulate district plans under the scenario that map drawers only have access to one out of the three versions of population counts (the original 2010 Census, the DAS-12.2, or the DAS-4.5).
Congressional district simulations were conducted with the Sequential Monte Carlo (SMC) redistricting sampler of \cite{mcca:imai:20}, while most of the state legislative district simulations use a Merge--Split MCMC sampler building from \cite{cart:etal:19,deford2019}.
Both of these sampling algorithms are implemented in the open-source software package \texttt{redist} \cite{kenn:etal:21}. The package allows simulating districts while imposing a population parity constraint so that all simulated maps are realistic.

Mirroring enacted maps, congressional district maps were sampled so that population deviations were at most 0.1 to 1 percent, and state legislative district population deviations were at most 5 to 10 percent, depending on the state.
We generated Monte Carlo samples until the standard diagnostics including the number of effective samples indicated accurate sampling and adequate sample diversity.
In the state legislative district simulations in South Carolina with over 100 districts and Mississippi with 52 districts, ensuring sampling diversity required running several chains of the Merge--Split algorithm in parallel, initiated from a sample generated from \cite{mcca:imai:20}.

In computing party's outcomes, we use precinct-level data from statewide elections from the Voting and Election Science Team (VEST) \cite{vest}.
The choice of statewide races avoids the variation in uncontested races and any incumbency effects in district races.
In Pennsylvania, we use the two-party vote share averaged across all statewide and Presidential races, 2004--2008, and adjust to match 2008 turnout levels.
In North Carolina we use the 2012 gubernatorial election at the precinct level.
In South Carolina we use the 2018 gubernatorial election and in Louisiana we use the 2019 Secretary of State election, each estimated at the voting tabulation district level, allocated based on 2010 Census block voting age population.
In Delaware, we use the precinct level returns from the 2020 presidential election.

The DAS-12.2 data yield precinct population counts that are roughly 1.0\% different from the original Census, and the DAS-4.5 data are about 1.9\% different.
For the average precinct, this amounts to a discrepancy of 18 people (for DAS-12.2) or 33 people (for DAS-4.5) moving across precinct boundaries.
Therefore, our main simulation results should be considered as a study of how such precinct-level differences propagate into noise at the district-level by exploring redistricting plans.

\begin{table}[t]
\centering
\begin{tabular}{llrrr}
\toprule
 \textbf{State} & \textbf{Office} & \textbf{Districts}  & \textbf{Precincts} & \textbf{Simulations}\\ \midrule
 Alabama & - & - & 1,992 & - \\
 Delaware &  State Senate & 21 & 434 & 10k \\
 Louisiana & State Senate & 39 & 3,668 & 35k\\
 Louisiana$^*$ & State House & 15 & 361 & 90k \\
 Mississippi & State Senate & 52 & 1,969 & 50k\\
 New York$^\dagger$ & School Board & 9 & 1,207 & 10k\\
 North Carolina & U.S. House & 13 & 2,692 & 200k\\
 Pennsylvania &  U.S. House & 18 & 9,256 & 10k\\
 South Carolina & U.S. House & 7 & 2,122 & 200k\\
 South Carolina & State House & 124 & 2,122 & 100k\\
 Utah & - & - & 2,337 & - \\
 Washington & - & - & 7,312 & -\\\bottomrule
\end{tabular}
\bigskip
\caption{States and districts studied. We compared the Census 2010, DAS-12.2, and DAS-4.5 datasets in seven states and three levels of elections. Simulations indicate the number of simulations for each of those three different comparison datasets. States that we only use for precinct-level modeling and not for redistricting simualtions are denoted by a dashed entry. \\$^*$Examines the Baton Rouge area.\\$^\dagger$Examines the East Ramapo school district, using Census blocks instead of voting precincts.}
\label{tab:simulations}
\end{table}

\hypertarget{population-parity}{%
\subsection{Population Parity}\label{population-parity}}

Perhaps the strongest constraint on modern redistricting is the requirement that districts be nearly equal in population.
Deviations in population between districts have the effect of diluting the power of voters in larger-population districts.
The importance of this principle stems from a series of Supreme Court cases in the 1960s, beginning with \textit{Gray v. Sanders} (1963), in which the court held that political equality comes via a standard known as \textit{One Person, One Vote}.
As for acceptable deviations from population equality, \emph{Wesberry v. Sanders} (1964) set the basic terms by holding that the Constitution requires that ``as nearly as is practicable one {[}person's{]} vote in a congressional election is to be worth as much as another's.'' Even minute differences in population parity across congressional districts must be justified, even when smaller than the expected error in decennial Census figures (\emph{Karcher v. Daggett} 1983).
For congressional districts, the majority of states thus balance population to within one person of perfect population parity \cite{ncsl:12}.
For state legislative districts, \textit{Reynolds v. Sims} (1964) held that they must be drawn to near population equality.
However, subsequent rulings stated that states may allow for small population deviations when seeking other legitimate interests (\emph{Mahan v. Howell} 1972; \emph{Gaffney v. Cummings} 1973).

When measuring population equality, states must rely on Census data, which was viewed as the most reliable source of population figures (\textit{Kirkpatrick v. Preisler} 1969).
We therefore empirically examine how the DAS affects the ability to draw redistricting maps that adhere to this equal population principle.
We simulate realistic maps for Pennsylvania Congressional districts and Louisiana State Senate districts based on the DAS-4.5 and DAS-12.2 data under various levels of population parity.
We then examine the degree to which the resulting maps satisfy the same population parity criteria using the 2010 Census data.

\hypertarget{partisan-effects}{%
\subsection{Partisan Effects}\label{partisan-effects}}

If changes in reported population in precincts affect the districts in which they are assigned to, this has implications for which parties win those districts.
While a change in population counts of about 1 percent may seem small, differences in vote counts of that magnitude can reverse some election outcomes.
Across the five U.S.
House elections during 2012 -- 2020, 25 races were decided by a margin of less than a percentage point between the Republican and Democratic party's vote shares.
And 228 state legislative races were decided by less than a percentage point from 2012--2016.

Partisan implications also raise the concern of gerrymandering, where political parties draw district boundaries to systematically favor their own voters.
Many uses of redistricting simulation in redistricting litigation have been over partisan gerrymanders, including \emph{Common Cause v. Lewis}, \emph{Rucho v. Common Cause}, \emph{Ohio A. Philip Randolph Institute v. Householder}, \emph{League of Women Voters of Michigan v. Benson}, and \emph{League of Women Voters v. Pennsylvania}.
To evaluate the impact of the DAS on the analysis of potential partisan gerrymanders we used the simulations from four states (Table \ref{tab:simulations}) compare the partisan attributes of the simulated plans from the three data sources.
We also analyze voting-related patterns in DAS-induced population count error at the precinct level, and connect these patterns to the statewide findings from the simulations.

\hypertarget{racial-effects}{%
\subsection{Racial Effects}\label{racial-effects}}

The Voting Rights Act of 1965, its subsequent amendments, and a series of Supreme Court cases all center race as an important feature of redistricting.
A large number of these cases focus on the creation of majority-minority districts (MMDs) (e.g.~\emph{Thornburg v. Gingles} 1986, \emph{Shaw v. Reno} 1993, \emph{Miller v. Johnson} 1995, \emph{Shelby County v. Holder} 2013).
First, we analyze whether the DAS data systematically undercounts or overcounts certain areas across racial lines.
In doing so, we focus on the consequences of the Bureau's decision to target accuracy to the majority racial group in a given area in their post-processing procedure \cite{census-das-targets}.

We also explore how DAS data can influence the creation of MMDs.
To do so, we empirically examine how using the DAS data to create MMDs differs from the same process undertaken using the 2010 Census data.
We simulate maps in the Louisiana State House using various levels of a constraint targeted to create MMDs and examine the degree to which maps generated using the Census and DAS data lead to different results at the precinct level.

\hypertarget{ecological-inference-and-voting-rights-analysis}{%
\subsection{Ecological Inference and Voting Rights Analysis}\label{ecological-inference-and-voting-rights-analysis}}

Researchers have developed methods to predict the race and ethnicity of individual voters using Census data.
Since \emph{Gingles}, voting rights cases have required evidence that an individual's race is highly correlated with candidate choice.
Statistical methods must therefore estimate this individual quantity from aggregate election results and aggregate demographic statistics \cite{good:53, king:rose:tann:04}.
A key input to these methods is accurate racial information on voters.
To produce this data, recent litigation has used Bayesian Improved Surname Geocoding (BISG) to impute race and ethnicity into a voter file \cite{fisc:frem:06,elli:etal:09,imai:khan:16}.
This methodology provides improved classification of the degree of racially polarized voting and racial segregation.

To understand how DAS data influence these analyses, we look at the effect of DAS data on the accuracy of the BISG across several states where race is recorded on the voter file so that we can assess classification accuracy.
We then re-examine the most recent racial gerrymandering case, \emph{NAACP, Spring Valley Branch et al.~v. East Ramapo School District} (2020), in which the BISG played a key role.
The federal Court of Appeals for the Second Circuit upheld the district court's ruling that the school board elections violated the Voting Rights Act.
We reanalyze this case using the DAS data and compare the results with those based on the 2010 Census data.

\hypertarget{biases}{%
\section{Racial and Partisan Undercounting Biases}\label{biases}}

The first step of the DAS pipeline is to add independent, symmetric Laplace or Geometric noise to counts in each Census table.
In addition, the DAS subsequently post-processes these noisy counts in order to ensure they are non-negative and consistent across tables and levels of the geographic hierarchy.
It is this post-processing step which can introduce substantial bias in population totals for certain areas, as we detail in this section.

\begin{figure}

{\centering \includegraphics[width=1\linewidth]{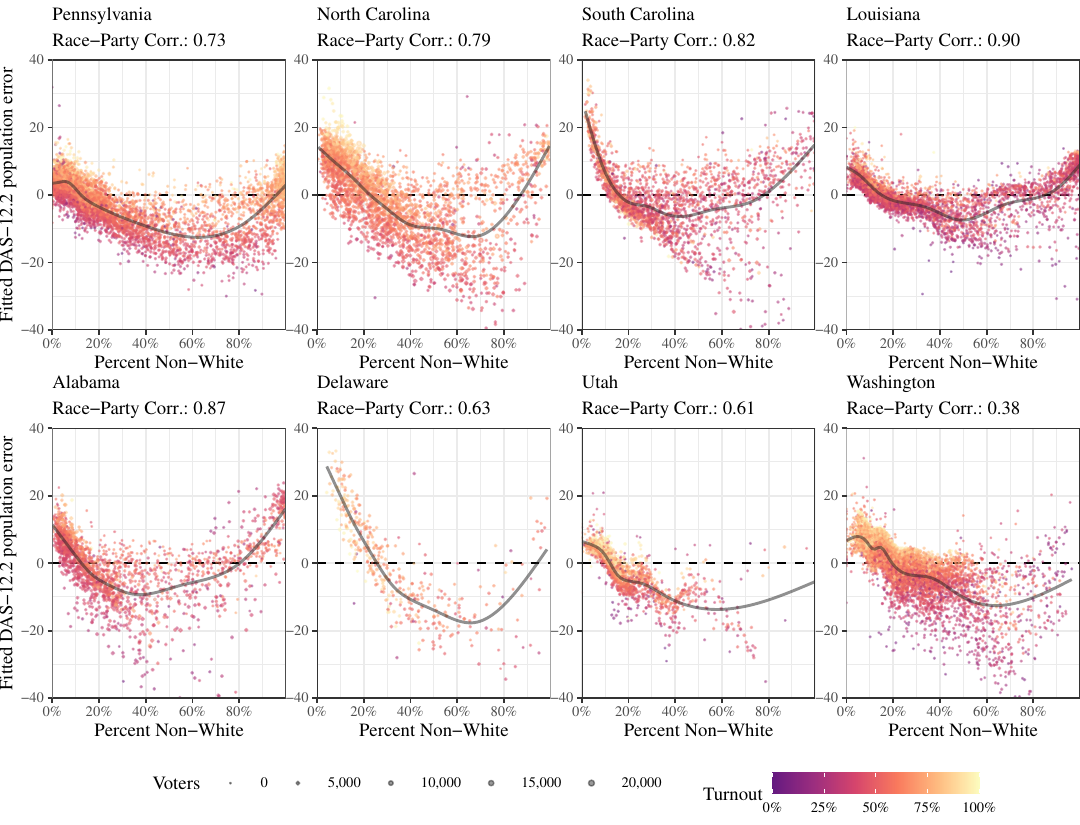} 

}

\caption{Model-smoothed error in precinct populations by the minority fraction of voter, with color indicating turnout. A GAM smooth is overlaid to show the mean error by minority share.}\label{fig:race-error}
\end{figure}

By the nature of the noise injection of the DAS, there is significant variation in the population error, even among similar precincts.
As a result, it is difficult to discern systematic patterns by observation alone.
We therefore fit a generalized additive model (GAM) to the precinct-level population errors, using various characteristics of the precinct.
Our predictors include the two-party Democratic vote share of elections in the precinct, turnout as a fraction of the Voting Age Population, log population density, the fraction of the population that is white, and the Herfindahl-Hirschman index of race as a measure of racial heterogeneity \cite{rosenbluth1955measures}.
The GAM regresses the difference in precinct population between the DAS-12.2 and the Census data on the following function of these predictors: \begin{align*}
{P}_{\text{DAS}, i} - {P}_{\text{Census}, i} \ = \ t(\text{Democratic}_i, \text{Turnout}_i,  \log(\text{Density}_i)) + s(\text{White}_i) + s(\text{HHI}_i) + \varepsilon_i,
\end{align*} where \(i\) indexes precincts or VTDs, \(P_{D, i}\) denotes \(i\)'s population as reported by data source \(D\), the function \(t\) indicates the smoothed tensor product cubic regression, the function \(s\) denotes thin-plate regression splines, and HHI denotes the Herfindahl-Hirschman index.
The model explains about 9--12 percent of the overall variance in population errors.

\begin{figure}

{\centering \includegraphics[width=1\linewidth]{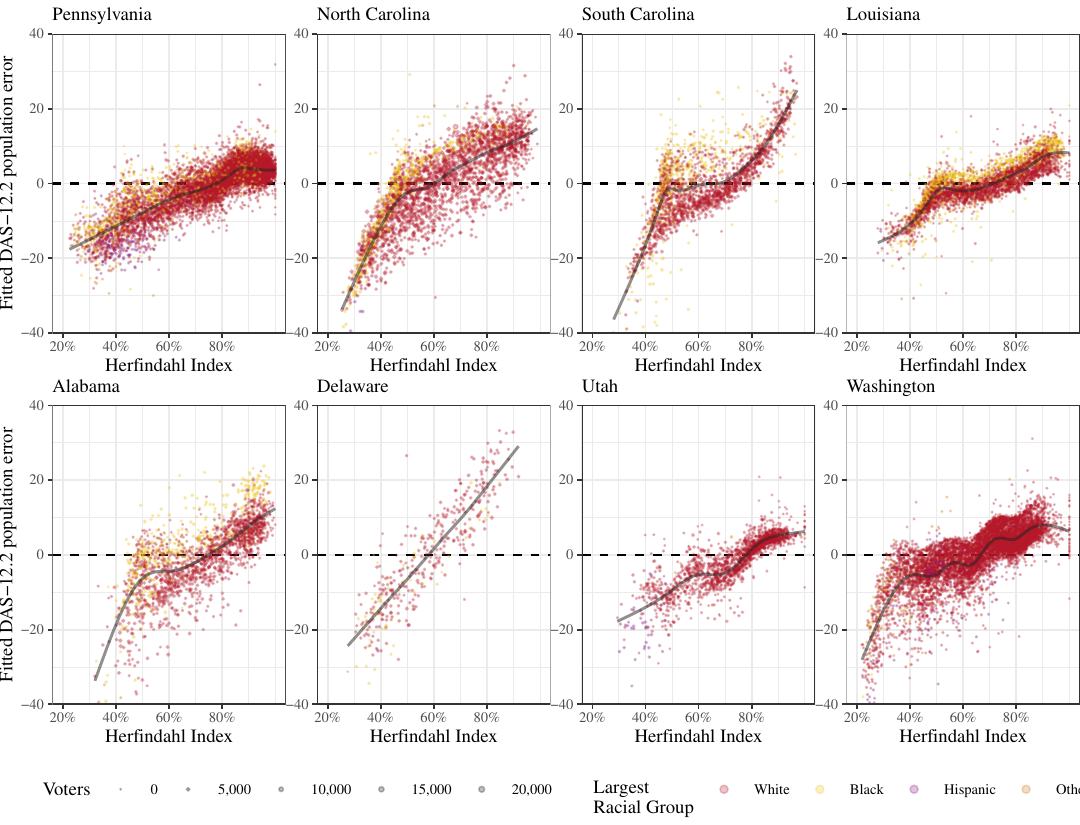} 

}

\caption{Model-smoothed error in precinct populations by the Herfindahl-Hirschman Index. A Herfindahl-Hirschman Index of 100 percent indicates that the precinct is composed of only one racial group.}\label{fig:herf-error}
\end{figure}

Figure \ref{fig:race-error} plots the fitted values from this model using deviations of the DAS-12.2 data against the minority fraction of the population in each VTD for eight states.
We chose to study a variety of states that are frequently studied in redistricting (Pennsylvania, North Carolina), the Rural South (South Carolina, Louisiana, Alabama), small states (Delaware), and heavily Republican (Utah) or Democratic (Washington) western states.
Consistent patterns emerge across these diverse states.
As indicated by U-shape patterns, mixed White/nonwhite precincts lose the most population relative to more homogeneous precincts.
Figure \ref{fig:herf-error} more clearly shows this pattern with racially homogeneous precincts.
We plot the error against the Herfindahl-Hirschman index, and find that the fitted error in estimated population steeply declines as the precinct becomes more racially diverse.

These patterns can be partially explained by the adopted DAS targets \cite{census-das-targets}, which prioritize accuracy for the largest racial group in a given area.
By doing so, the DAS procedure appears to undercount heterogeneous areas where the population differences between racial groups are relatively small.
In highly heterogeneous precincts, the accuracy guarantees are thus much lower.
As precincts are the building blocks of political districts, our results demonstrate that precincts that are heterogeneous along racial and partisan lines would lose electoral power under the DAS.
In aggregate, the movement of population from heterogeneous to homogeneous precincts would tend to increase the apparent spatial segregation by race.

\begin{figure}

{\centering \includegraphics[width=1\linewidth]{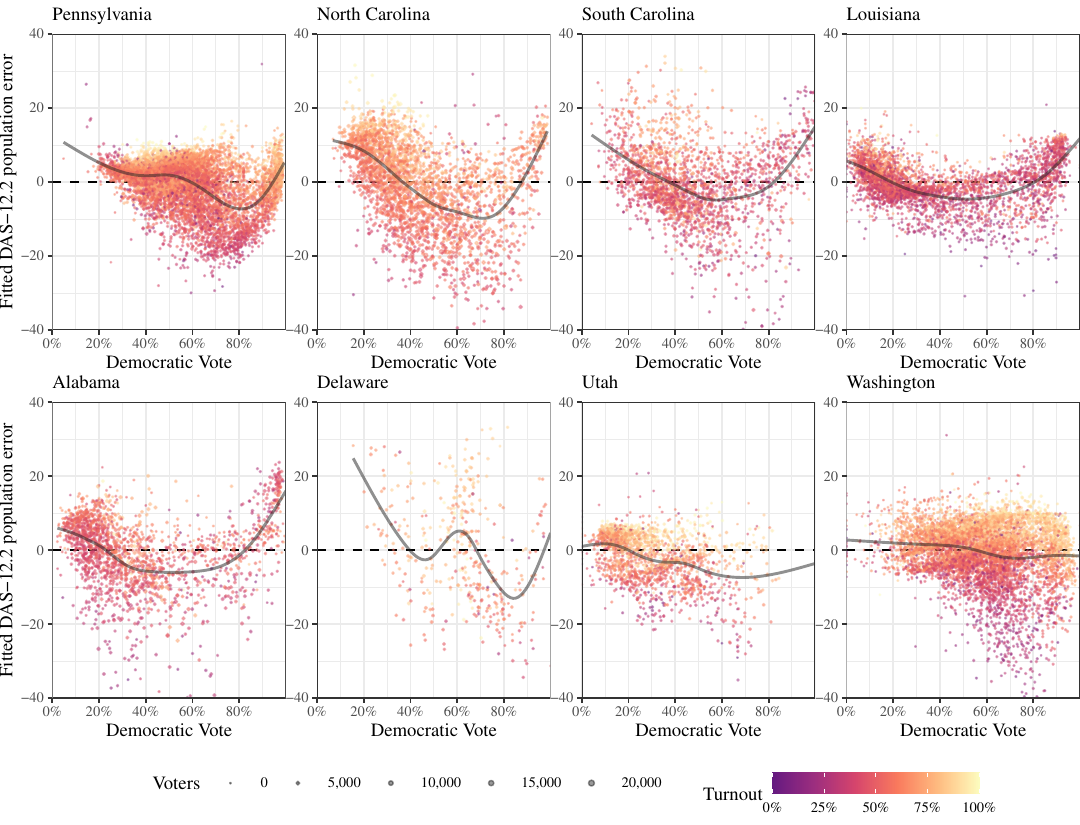} 

}

\caption{Model-smoothed error in precinct populations by Democratic two-party vote share, with color indicating turnout. A GAM smooth is overlaid to show the mean error by Democratic share.}\label{fig:partisan-error}
\end{figure}

Figure \ref{fig:partisan-error} shows similar patterns of undercounting biases along a partisan dimension.
Moderately Democratic precincts are on average assigned less population under the DAS than the actual 2010 Census.
Further, higher-turnout precincts are on average assigned more population under the DAS than they should otherwise have.
These effects are on the order of 5--15 voters per precinct, on average, though some are larger.\footnote{Not shown is the equivalent figure for the DAS-4.5 data, which displayed an identical pattern but with roughly double the magnitude of fitted error.
  This is available as Figure \ref{fig:a-partisan-error} in Appendix B.}
Aggregated across the hundreds of precincts that comprise the average district, however, the errors may become substantial, as we discuss in more detail in the sections below.
In the 70 congressional districts in the states we examine statwide, the average district's population changes by 308 people when measured with DAS-12.2 counts.
But, in two Pennsylvania congressional districts in and around Philadelphia, the population changes by 2,151 people on average.
This measured difference under the DAS is orders of magnitude larger than the difference under block population numbers released in 2010.

It is difficult to know exactly how these partisan and racial biases arise without more detail on the DAS post-processing system and parameters.
Regardless, the presence of differential bias in the precinct populations according to partisanship, turnout, and racial diversity is concerning.

\hypertarget{population-parity-in-redistricting}{%
\section{Population Parity in Redistricting}\label{population-parity-in-redistricting}}

Deviation from population parity across \(n_d\) districts is generally defined as

\begin{align*}
  \textrm{deviation from parity} = \max_{1 \le k \le n_d} \frac{|P_k-\overline{P}|}{\overline{P}},
\end{align*}

where \(P_k\) denotes the population of district \(k\) and \(\overline P\) denotes the target district population.
In other words, we track the percent difference in the district population \(P_k\) from the average district size \(\overline{P}\), and report the maximum deviation.
Our redistricting simulations generate plans that do not exceed a user-specified tolerance.
After generating these plans, we then re-evaluate the deviation from parity using the precinct populations from the three data sources.

We find that the noise introduced by the DAS prevents the drawing of equal-population maps with commonly-used population deviation thresholds.
Because only one dataset will be available in practice, redistricting practitioners who attempt to create equal-population districts with DAS data should expect the actual deviation from parity to be significantly larger than what they can observe in their data.
Due to the asymmetric post-processing within the DAS algorithm, there is no clear way to improve estimates.
We find that this problem is more acute in state legislative districts, where there are more districts and each district is composed of fewer precincts.

\hypertarget{congressional-districts-in-pennsylvania}{%
\subsection{Congressional Districts in Pennsylvania}\label{congressional-districts-in-pennsylvania}}

\begin{figure}

{\centering \includegraphics[width=1\linewidth]{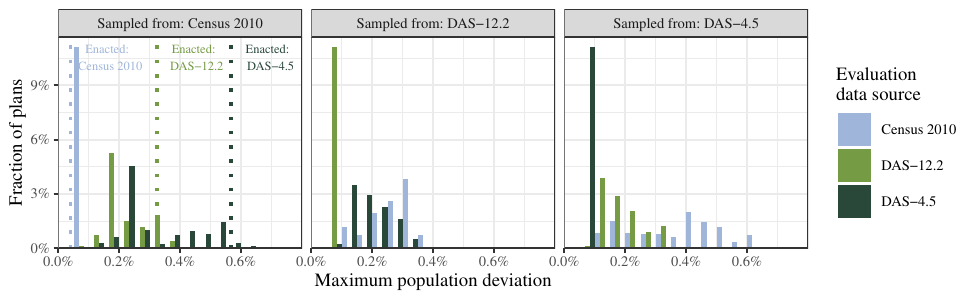} 

}

\caption{Maximum deviation from population parity among Pennsylvania redistricting plans simulated from the three data sources. All plans were sampled with a population constraint of 0.1 percent, corresponding to the deviation measured from the Census 2010 precinct data, and marked with the dashed line. Deviation from parity was then evaluated using the three versions of population data.}\label{fig:pa-parity}
\end{figure}

Figure \ref{fig:pa-parity} shows the maximum deviation from population parity for the 30,000 simulated redistricting plans in Pennsylvania, when evaluated according to the three different data sources.\footnote{10,000 plans were simulated from each data source, with every plan satisfying a 0.1\% population parity constraint.
  The simulation algorithm also ensured that no more than 17 counties were split across the entire state, reflecting the requirement in Pennsylvania that district boundaries align with the boundaries of political subdivisions to the greatest extent possible.}
Consistently, plans that were generated under one set of population data and drawn to have a maximum deviation of no more than 0.1\% had much larger deviations when measured under a different set of population data.
For example, of the 10,000 maps simulated using the DAS-12.2 data (the middle panel of the figure), 9,915 exceeded the maximum population deviation threshold, according to the 2010 Census data.
While nearly every plan failed to meet the population deviation threshold, the exact amount of error varied significantly across the simulation set.
As a result, redistricting practitioners who attempt to create equal-population districts according to similar thresholds can expect the actual deviation from parity to be significantly larger but of unknown magnitude.

\hypertarget{state-legislative-districts-in-louisiana}{%
\subsection{State Legislative Districts in Louisiana}\label{state-legislative-districts-in-louisiana}}

\begin{figure}

{\centering \includegraphics[width=1\linewidth]{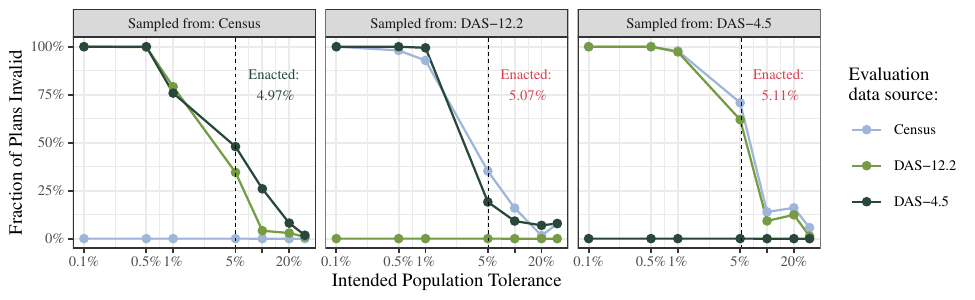} 

}

\caption{Fraction of Louisiana State Senate plans simulated under one data source with a population parity constraint which are invalid when measured under another. The horizontal axis shows the tolerance constraint for the original simulation on the log10 scale. The vertical axis shows the percent of plans that exceed the intended tolerance according to the evaluation data.  The dashed line shows the maximum deviation from parity of the enacted 2010 map.}\label{fig:la-parity}
\end{figure}

We expect smaller districts such as state legislative districts to be more prone to discrepancies in population parity.
For example, the average Louisiana Congressional district comprises about 600 precincts, but a State Senate district comprises about 90 and a State House district only 35.
Therefore, deviations due to DAS are more likely to result in larger percent deviations from the average.
To test this, we compared 35,000 Louisiana State Senate plans generated from each of the three data sources (105,000 in total) and population parity constraints ranging from 0.1\% to 20\%, measuring the plans' population deviation against the three different data sources.\footnote{5,000 plans were simulated for each data source/population parity pair.}
Figure \ref{fig:la-parity} plots the results of this comparison.

As expected, we see complete acceptance for plans measured with the dataset from which they were generated.
However, plans generated under one dataset can exceed the population threshold under another.
Specifically, plans generated under DAS data can be very likely to be invalid when evaluated using the true Census data (see the middle and right panels of the figure).
The rate of invalid plans grows as the tolerance becomes more precise.

Also noteworthy is the fact that even at the population parity tolerances as generous as 1.0\%, all generated plans are invalid in some cases.
Compared to Pennsylvania congressional districts, with a parity tolerance of 0.1\%, simulated districts for the Louisiana State Senate fail to meet the cutoffs much more often, as the DAS-added noise is relatively larger at smaller scales.

\hypertarget{partisan-effects-on-redistricting}{%
\section{Partisan Effects on Redistricting}\label{partisan-effects-on-redistricting}}

How does the systematic undercounting and overcounting of precinct-level populations in the DAS data, affect the conclusions we draw about the partisan and racial biases of legislative redistricting plans?
We find that the precinct-level biases may aggregate in unexpected ways, and add to possible enumeration errors, making it more difficult to identify gerrymanders.

\begin{figure}

{\centering \includegraphics[width=1\linewidth]{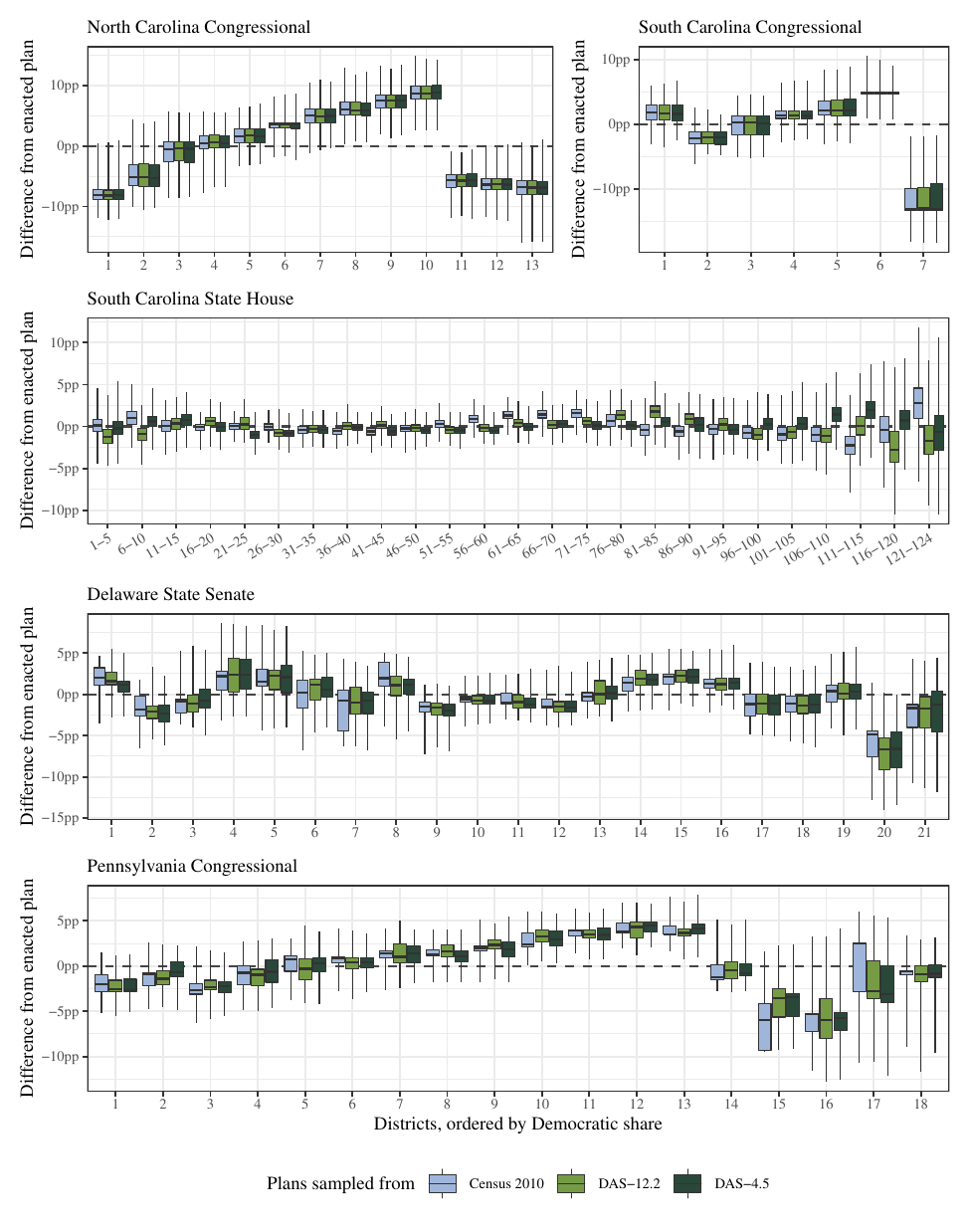} 

}

\caption{The distribution of the simulated Democratic vote share relative to the enacted redistricting plan, comparing the original Census with DAS versions. Districts are ordered by the level of the Democratic vote in each simulation.  Districts are grouped for the South Carolina State House due to the large number of districts. Whiskers extend to cover the full range of the simulated data.}\label{fig:partisan-boxplots}
\end{figure}

We first assess the impact of DAS data in identifying partisan packing and cracking, following a common approach in redistricting analysis.
Practitioners and researchers compare enacted plans against a distribution of election results from each simulated plan, for example using actual precinct-level vote counts in statewide districts and adding them up to each simulated district.
Plans which are partisan gerrymanders stand out from the simulated ensemble as yielding more seats for one party over the other.
The argument that ensemble analysis is sufficient for this purpose has been used in various academic contexts, including \cite{chen:rodd:15, hers:etal:20}, and litigation contexts, most recently \cite{poli-sci-amicus:19, pegd:rodd:wang:19}.

The results from non-simulation analysis (Figure \ref{fig:partisan-error}) already suggests that the DAS-induced noise may not cancel out if diverse areas are spatially clustered.
The systematic patterns at the district-level clearly depends on the spatial adjacency of diverse and homogeneous precincts.
Simulations can evaluate these implications beyond particular enacted plans.

We find that across tens of thousands of simulated plans, the DAS leads to unpredictable differences in the distribution state-level party outcomes under the three data sources.
Figure \ref{fig:partisan-boxplots} summarizes the differences in simulations with a plot commonly used for identifying outlier plans.
To compare district-level partisan outcomes across simulations, we sort districts in ascending order by Democratic two-party vote share in each simulation, so that, district number 1 in North Carolina is always the most Republican district in each simulated plan and district number 13 is the most Democratic district in the same plan.
For each ordered district we subtract off the enacted plan's Democratic vote from all the simulations and plot the differences in a boxplot (whiskers extend the entire range of simulated data).
A boxplot is completely above zero indicates that the enacted plan had fewer Democratic voters in that district than would be expected under a partisan-neutral baseline---in other words, that the district cracked Democratic voters.

The patterns in Figure \ref{fig:partisan-boxplots} are concerning.
While in some cases, such as the congressional districts in North and South Carolina, there are no discernible differences across the three data sources, for others, the differences can be substantial.
For the South Carolina State House, a pattern of cracking in the 61st--75th most Democratic districts under the Census 2010 data disappears under the DAS-protected data.
Even more concerning, evidence of packing in the 111th--115th most Democratic state legislative districts under the Census 2010 data is reversed under the DAS-4.5 data.
In Pennsylvania, results are relatively stable across data sources for the relative Republican districts 1--14, but display significant differences for the most Democratic districts (15 and 17), with median discrepancies moving as much as five percentage points.

Given that redistricting litigation often must focus on a single district or set of districts \cite{issacharoff1997standing}, discrepancies of this magnitude at the district level could change the conclusions regarding the presence or absence of a partisan gerrymander.
The fact that the presence and magnitude of the discrepancies are not consistent even within the same state is also worrying, as it complicates efforts to take into account these potential biases in research and decision making.

\hypertarget{racial-effects-on-redistricting}{%
\section{Racial Effects on Redistricting}\label{racial-effects-on-redistricting}}

\begin{figure}

{\centering \includegraphics[width=1\linewidth]{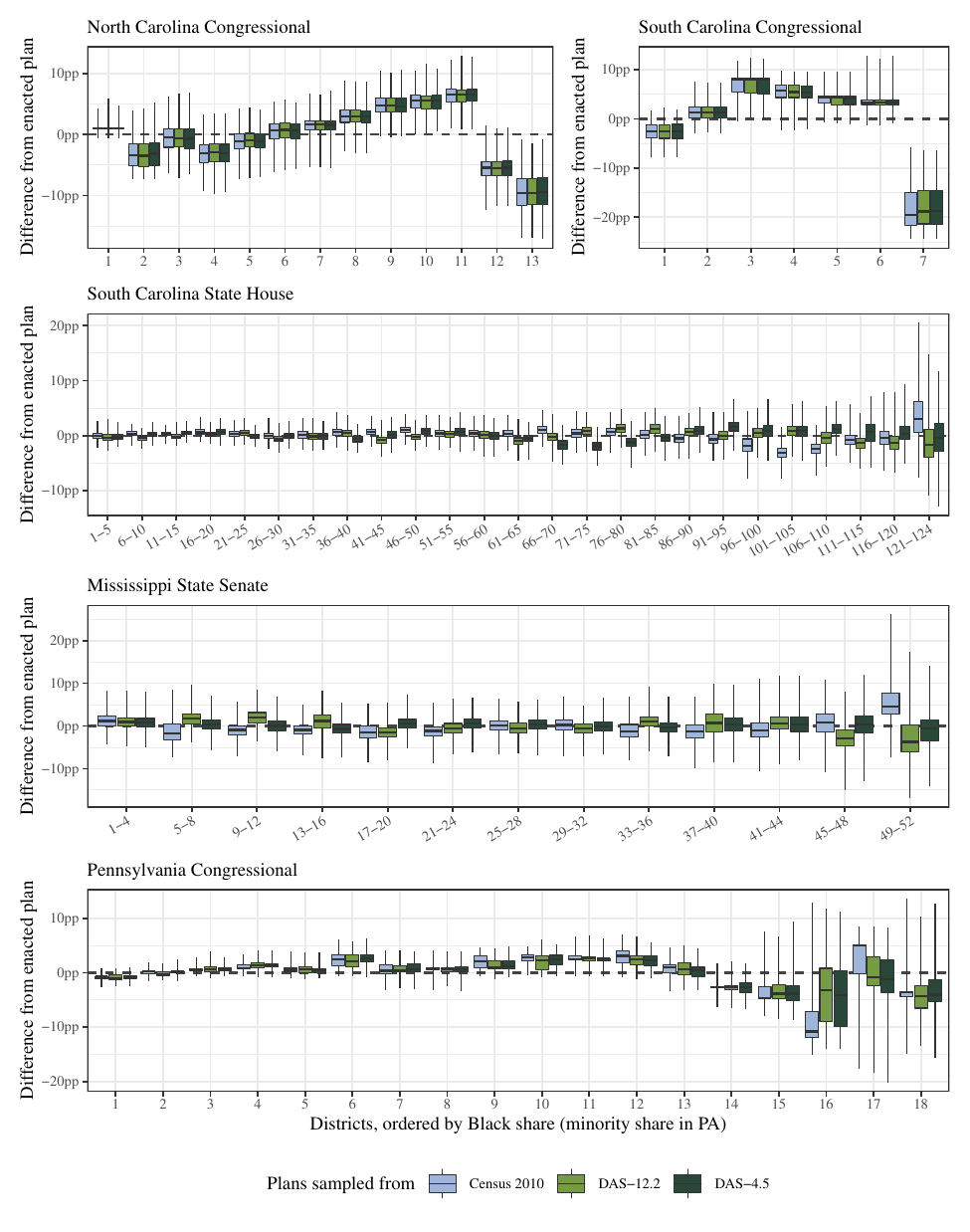} 

}

\caption{The distribution of the simulated Black population share relative to the enacted redistricting plan, comparing the original Census with DAS versions. Districts are in numbered in ascending order of the Black or minority population share in each simulation. Districts are grouped for the South Carolina State House and Mississippi State Senate due to the large number of districts. Whiskers extend to cover the full range of the data.}\label{fig:race-boxplots}
\end{figure}

We perform a similar analysis of discrepancies by comparing the distribution of the minority population in each district.
Figure \ref{fig:race-boxplots} orders the districts from the same simulation as Figure \ref{fig:partisan-boxplots} by their Black population share, and compares those shares with the ordered districts in the enacted plan.

As with partisan error, there are inconsistent patterns across states and district sizes.
The racial makeup of congressional districts in North and South Carolina do not appear to be affected.
For the South Carolina State House, patterns of cracking in the districts with the largest racial minorities (districts ordered 121-124) under the Census 2010 data disappear or are even reversed under the DAS-12.2 and DAS-4.5 data.
In districts ranked to be the 96th to 110th most Black, patterns of packing are reversed in the DAS data.
Similarly, in the Mississippi State Senate, evidence of packing in the most Black districts (ordered 49-52) becomes evidence of cracking under the DAS-4.5 data.
In Pennsylvania's 18 Congressional districts, patterns are generally stable across the data sources for the 14 most White congressional districts, but display significant differences for the heavily non-White districts ordered 16 and 17, with median discrepancies moving as much as seven percentage points.

\begin{figure}

{\centering \includegraphics[width=1\linewidth]{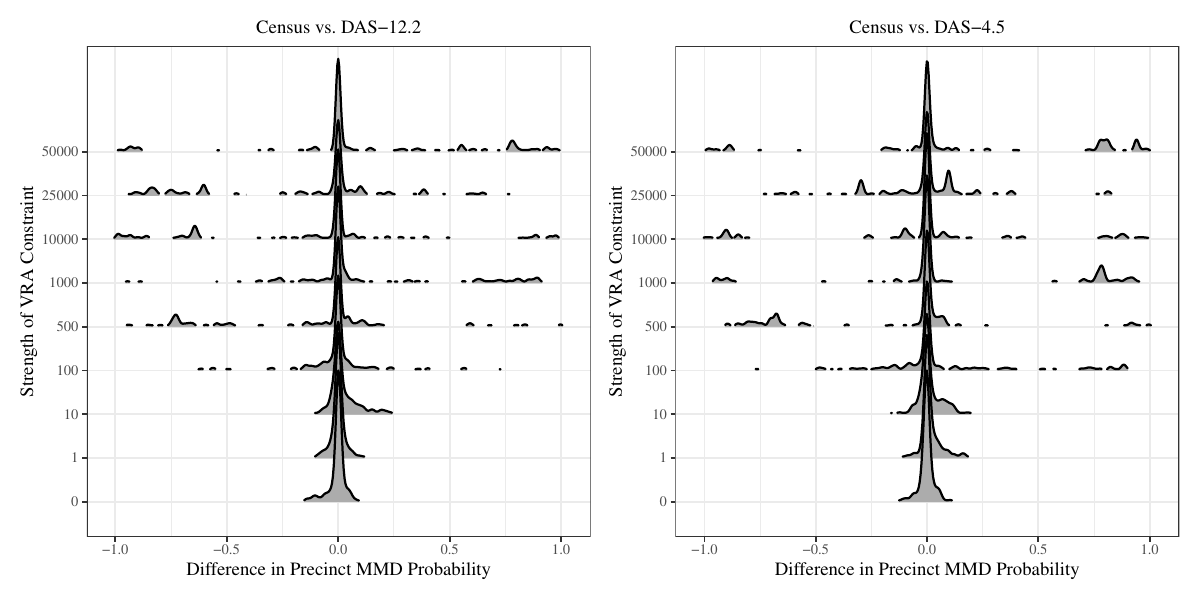} 

}

\caption{The left (DAS-12.2) and right (DAS-4.5) columns show that precinct-level assignments can differ substantially in simulated districts generated using the 2010 Census and DAS data. Here, the calculated probability of being assigned to a majority-minority district can be much higher or lower for individual precincts, and these differences grow as a constraint encouraging the formation of MMDs is strengthened.}\label{fig:ebr-race}
\end{figure}

These district-level findings may still mask the variability around which individual precincts are included in majority minority districts.
In the left and right columns of Figure \ref{fig:ebr-race}, we show the results of 10,000 simulations of the Merge--Split Markov chain Monte Carlo (MCMC) sampler with various levels of a Voting Rights Act (VRA) constraint.
This constraint, which we did not apply in the previous sections, encourages the formation of majority-minority districts.
We then calculate the probability that each precinct is assigned to a majority-minority district (as defined by Black population).
Finally, we calculate the difference between these probabilities for the Census versus DAS-12.2 and Census versus DAS-4.5.

With no VRA constraint (corresponding to the VRA strength of zero on the \(y\)-axis), each precinct has similar probabilities of being in a MMD, regardless of the dataset used.
However, as the strength of this constraint increases (making the algorithm search for MMDs more aggressively), we see that the noise introduced to the DAS data systematically alters the district membership of individual precincts.
A precinct with a value of \(1\) or \(-1\) in the left and right columns of Figure \ref{fig:ebr-race} indicates that those precincts are never in a MMD under one dataset but are always in a MMD when the same mapmaking process is done with a different dataset.
This means that essentially random changes in how Census data is reported could meaningfully influence the political representation of voters in particular precincts.

\hypertarget{ecological-inference-and-voting-rights-analysis-1}{%
\section{Ecological Inference and Voting Rights Analysis}\label{ecological-inference-and-voting-rights-analysis-1}}

Inferring the racial and ethnic composition of potential voters and their candidate choice is a key element of voting rights analysis in redistricting.
Recent court cases have relied on Bayesian Improved Surname Geocoding (BISG) to predict the race and ethnicity of individual voters in a voter file \cite{fisc:frem:06,elli:etal:09,imai:khan:16}.
This methodology combines the names and addresses of registered voters with block-level racial composition data from the Census.

We first examine how the accuracy of prediction changes between the DAS and original Census data.
While differential privacy should not prevent statistical prediction, race is the most sensitive information included in the P.L.94-171 data release.
As such, it is of interest to examine whether the DAS will degrade the prediction accuracy of individual race and ethnicity.
We follow up on this analysis by revisiting a recent Section 2 court case about the East Ramapo school board election and investigate whether this change in racial prediction alters the conclusions of the racial redistricting analysis.

\hypertarget{prediction-of-individual-voters-race-and-ethnicity}{%
\subsection{Prediction of Individual Voters' Race and Ethnicity}\label{prediction-of-individual-voters-race-and-ethnicity}}

We first compare the accuracy of predicting individual voters' race and ethnicity using the original 2010 Census data, the DAS-12.2 data, and the DAS-4.5 data.
To obtain the benchmark, we use the North Carolina voter file acquired in February 2021.\footnote{We obtain the voter files used in this paper through L2, Inc., which is a leading national nonpartisan firm that supplies voter data and related technology.}
In several southern states including North Carolina,\footnote{The other states are Alabama, Florida, Georgia, Louisiana, and South Carolina.} the voter files contain the self-reported race of each registered voter
. This information can then be used to assess the accuracy of the BISG prediction methodology
.

Our approach follows that of \cite{imai:khan:16}.
We denote by \(E_i\) the race and ethnicity of voter \(i\), \(N_i\) as the surname of voter \(i\), and \(G_i\) as the geography in which voter \(i\) resides.
For each choice of race and ethnicity \(e \in \mathcal{E} = \{\)White, Black, Hispanic, Asian, Other\(\}\), Bayes' rule implies \[\Pr(E_i = e \mid N_i = n, G_i = g) \ = \ \frac{\Pr(N_i = n \mid E_i = e) \Pr(E_i = e \mid G_i = g)}{\sum_{e' \in \mathcal{E}} \Pr(N_i = n \mid E_i = e') \Pr(E_i = e' \mid G_i = g)},\]where we have assumed the conditional independence between the surname of a voter and their geolocation within each racial category, i.e., \(N_i \perp \!\!\! \perp G_i \mid E_i\).

In the presence of multiple names---e.g.
first name \(f\), middle name \(m\), and surname \(s\)---we make the further conditional independence assumption \cite{voic:18} \begin{align*}
\Pr(N_i = &\{f, m, s\} \mid E_i = e) \\
&= \ \Pr(F_i = f \mid E_i = e) \Pr(M_i = m \mid E_i = e) \Pr(S_i = s \mid E_i = e),
\end{align*} where \(F_i, M_i,\) and \(S_i\) represent individual \(i\)'s first, middle, and surnames respectively.

We compare estimates by changing the data source from which the geographic prior, \(\Pr(E_i = e \mid G_i = g)\), is estimated, from the 2010 Census to each of the two DAS datasets.
Estimates of the other race prediction probabilities are obtained by merging three sources: the 2010 Census surname list \cite{censusSurnameList}, the Spanish surname list from the Census, and the voter files from six states in the U.S.
South, where state governments collect racial and ethnic data about registered voters for Voting Rights Act compliance.
The middle and first name probabilities are derived exclusively from the voter files.

We evaluate the accuracy of the BISG methodology on approximately 5.8 million registered voters included in the North Carolina February 2021 voter file.
Among them, approximately 70\% are White and 22.5\% are Black, with smaller contingents of Hispanic (3.4\%), Asian (1.5\%), and Other (2.4\%) voters.

\begin{figure}

{\centering \includegraphics[width=5.5in]{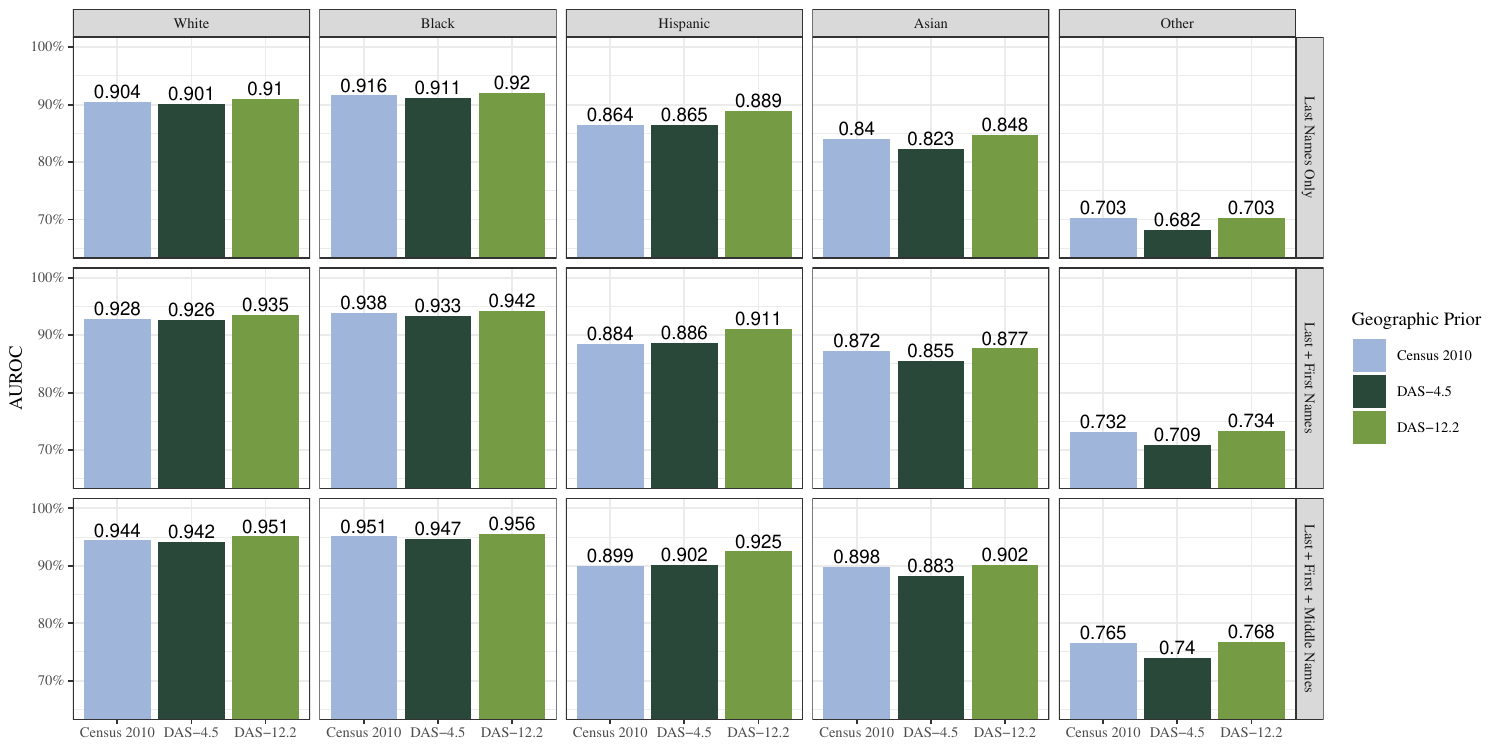} 

}

\caption{Area under the Receiver Operating Characteristic Curve (AUROC) percentage values for the prediction of individual voter's race and ethnicity using North Carolina voter file. Bars represent AUROC with geographic priors given by each of three datasets: 2010 Census, DAS-4.5, and DAS-12.2.}\label{fig:nc-roc-results}
\end{figure}

Figure \ref{fig:nc-roc-results} summarizes the accuracy of the race prediction with the area under the Receiver Operating Characteristic curve (AUROC), which ranges from 0 (perfect misclassification) to 1 (perfect classification).
Across all racial and ethnic groups except Hispanics, we find the same surprising pattern: relative to the 2010 Census data, the DAS-12.2 data yield a small improvement in prediction performance while the DAS-4.5 data give a slight degradation.
Among Hispanics, both forms of DAS-protected data result in slightly improved predictions over the original Census data.

The strong performance of the DAS data in this setting is counter-intuitive.
It is possible that the noise added to the underlying data has somehow mirrored the true patterns of population shift from 2010 to 2021; or that this noise makes the DAS-12.2 data more reflective of the voter population relative to the voting-age population.
Additionally, the DAS may degrade or attenuate individual probabilities without having a significant impact on the overall ability to classify, something that AUROC is not designed to measure \cite{lobo2008auc}.

Results are substantively similar if we consider the classification error, under the heuristic that we assign each individual to the racial and ethnic group with the highest posterior probability.
Using the true Census data to establish geographic priors, we achieve posterior misclassification rates of 15.1\%, 12.1\%, and 10.0\% when using the last name; last name and first name; and last, first, and middle names for prediction, respectively.
The analogous misclassification rates are slightly higher for the DAS-4.5 priors---15.6\%, 12.5\%, and 10.3\%---but the same or slightly lower for the DAS-12.2 priors: 15.1\%, 12.0\%, and 9.9\%.

Our analysis shows that across three main racial and ethnic groups, the predictions based on the DAS data appear to be as accurate as those based on the 2010 Census data.
The finding suggests that although the new DAS methodology may protect \emph{differential} privacy, it may not prevent accurate prediction of sensitive attributes when compared to the swapping methodology used in the 2010 Census.

\hypertarget{ecological-inference-in-the-voting-rights-analysis}{%
\subsection{Ecological Inference in the Voting Rights Analysis}\label{ecological-inference-in-the-voting-rights-analysis}}

The BISG methodology played a central role in the most recent court case regarding Section 2 of the Voting Rights Act, \emph{NAACP of Spring Valley v. East Ramapo Central School District} (2020).
The East Ramapo Central School District (ERCSD) nine-member school board was elected using at-large elections.
This often led to an all White school board, despite 35\% of the voter eligible population being Black or Hispanic.
Yet, within the district, nearly all White school children attend private yeshivas, whereas nearly all Black and Hispanic children attend the ERCSD public schools.
As a result of this case, the district moved to a ward system.

We re-examine the remedy of this case, focusing on effective majority-minority districts (MMDs) based on a voter file with individual race and ethnicity imputed using the DAS-12.2 and Census 2010 data.
To approximate the data used by an expert witness who testified in the court case, we obtain the New York voter file (as of November 16, 2020) from the state Board of Elections.
We subset the voters to active voters with addresses in Rockland County, where ERCSD is located.
Using the R package \texttt{censusxy}, which interfaces with the Census Bureau's batch geocoder, we match each voter to a block and subset the voters to those who live within the geographic bounds of ERCSD \cite{censusxy, pren:fox:21}.
This leaves 58,253 voters, for whom we impute races using the same machinery behind the R package \texttt{wru} \cite{wru}, as described in \cite{imai:khan:16}.
This process closely mimics the one used in the original case.

\begin{figure}

{\centering \includegraphics[width=0.9\linewidth]{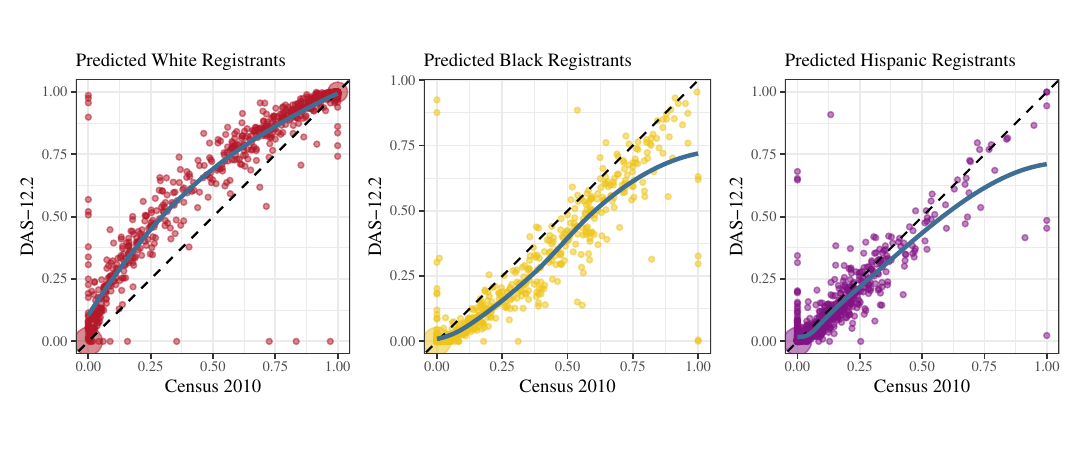} 

}

\caption{Imputed Racial Registrants by Census Blocks. The x-axis represents the percent of a group, as measured by the most likely race from racial imputation using the Census 2010 data. The y-axis represents the corresponding imputation using the DAS-12.2 data.}\label{fig:er-race-scatter}
\end{figure}

We examine how the predictions of individual race and ethnicity based on the 2010 Census and DAS-12.2 data result in different redistricting outcomes.
Figure \ref{fig:er-race-scatter} compares these two predictions using the proportions of (predicted) Whites, Black, and Hispanic registered voters for each Census block.
We find that the predictions based on the DAS-12.2 tend to produce blocks with more White registered voters than those based on the original Census data.
As a consequence, the predicted proportions of Black and Hispanic registrants are much smaller, especially in the blocks where they form a majority group.

The precise reason for these biases is unclear.
The DAS tends to introduce more error for minority groups than for White voters, and even more error for voters who are in a minority group for their Census block, which is more common for minority voters as well.
This additional noise, when carried through a nonlinear transformation such as the Bayes' rule calculation for racial imputation, may introduce some bias.
In addition, the large bias for White and Black voters relative to Hispanic voters suggests that the similarity of surnames between the White and Black populations, compared to the Hispanic population, may also be a factor.
Regardless, it is clear that the DAS-injected noise differentially biases voter race imputations at the block level.
This pattern may not always yield greater inaccuracies when aggregated to the statewide level---as seen in the prior section---but it is especially prevalent within the ERCSD.

\begin{table}

\caption{\label{tab:er-pair}East Ramapo MMDs under Census 2010 and DAS-12.2 data. The noise introduced in the DAS-12.2 leads us to undercount the number of majority minority districts in many plans, but never to overcount them.}
\centering
\begin{tabular}[t]{c>{\centering\arraybackslash}p{1cm}>{\centering\arraybackslash}p{1cm}>{\centering\arraybackslash}p{1cm}>{\centering\arraybackslash}p{1cm}r}
\toprule
\multicolumn{1}{c}{ } & \multicolumn{4}{c}{Number of MMDs from DAS-12.2} & \multicolumn{1}{c}{ } \\
\cmidrule(l{3pt}r{3pt}){2-5}
Census 2010 & 0 & 1 & 2 & 3 & Plans\\
\midrule
0 & \textbf{100\%} & 0 & 0 & 0 & 50\\
1 & 10 & \textbf{90} & 0 & 0 & 3,103\\
2 & 2 & 56 & \textbf{42} & 0 & 6,774\\
3 & 0 & 77 & 22 & \textbf{0} & 73\\
\bottomrule
\multicolumn{6}{l}{\textsuperscript{} Note: Percentages add to 100\% by row.}\\
\end{tabular}
\end{table}

We next investigate whether these systematic differences in racial prediction lead to different redistricting outcomes.
Specifically, we simulate 10,000 redistricting plans using DAS-12.2 population and a 5\% population parity tolerance.
We find that the systematic differences in racial prediction identified above results in the underestimation of the number of MMDs in these plans.
As in the original court case, an MMD is defined as a district, in which more than 50\% of its registered voters are either Black or Hispanic.
Table \ref{tab:er-pair} clearly shows that the number of MMDs based on the DAS-12.2 data never exceeds that based on the 2010 Census for all simulated plans.
For example, among 6,774 plans that are estimated to yield 2 MMDs according to the Census data, 56\% of them are predicted to have only 1 MMD.

While one should not extrapolate from this single case study, our analysis implies that in small electoral districts such as those of school board elections, the DAS can generate bias that may favor one racial group over another.
Although the number of MMDs is underestimated under the DAS data in this case, the direction and magnitude of racial effects are difficult to predict, as they depend on how the choice of tuning parameters in the DAS algorithm interact with a number of geographical and other factors.

\hypertarget{conclusion}{%
\section{Conclusion}\label{conclusion}}

The Census Bureau is faced with the difficult task of balancing its statutory mandate for protecting respondent privacy with the accuracy of its reported count.
The Bureau's Disclosure Avoidance System (DAS) clearly reflects this critical trade-off.
The DAS relies on differential privacy and adding random noise to the raw Census counts while it also uses a complex post-processing procedure to avoid negative counts and maintain consistency of published population counts across several levels of geographical units.

In this paper, we studied the impacts of the DAS on redistricting analyses using the most recent Demonstration data released in April 2021.
We considered a likely scenario, in which map drawers and analysts treat the noise-injected DAS data ``as-is,'' without performing any additional accounting for the DAS noise.
We find that the DAS has profound effects on standard redistricting analyses and procedures.
Despite the efforts of the Bureau to minimize error, we find that the added noise artificially shifts population counts from racially heterogeneous and mixed-partisanship areas to more homogeneous areas.
These non-random local errors can aggregate into substantively large and unpredictable biases at district levels, especially for small districts.
Fixing these systematic biases is of fundamental importance, as they will have partisan and racial impacts on the upcoming redistricting.

We also find that the added noise makes it impossible to follow the principle of \emph{One Person, One Vote} as it is currently interpreted by courts and policymakers.
Specifically, the principle requires states to minimize population differences between districts as much as possible.
Given the magnitude of population errors introduced by the DAS, our analysis shows that current practices of redistricting will make it difficult, and in some cases effectively impossible, to meet these existing standards.
In the near future, courts may decide to treat this error as-is or loosen the bounds on these standards.
Such a move will change the precedents and fundamentally shift our understanding of congressional redistricting.
It may also impact the partisan balance of power.

The complex nature of the DAS post-processing procedure masks the ultimate source of these biases.
Our findings suggest that they are likely a combination of several factors.
First, the bias against heterogeneous areas suggest could be driven by the Bureau's decision to target accuracy to the population count of the majority racial group in a given area \cite{census-das-targets}.
Second, the choice to not prioritize accuracy at the block level leads to an additive effect in many cases.
Our precinct-level population tabulations reveal around a 1\% average deviation in the DAS-12.2 data compared to the 2010 Census data, and these errors do not always cancel out.
Ensuring that population is accurate at this off-spine geography would help minimize population deviations among the majority of states which rely on these geographies to draw and evaluate their districts.

One general strength of the differential privacy framework is that the noise generation mechanism is known.
However, the asymmetric and deterministic nature of this post-processing procedure of the DAS makes such a proper statistical adjustment difficult for many commonly used models and near-impossible for others.
For this reason, many analysts are likely to treat the DAS-protected data as the basis for evaluating districts, as they have done in the past.
One possible approach is for the Bureau to additionally release the noisy DAS data without post-processing so that analysts can use it for their statistical analysis.
This will not solve the problems in map drawing, but would allow researchers to properly calibrate uncertainty at least for some analyses when evaluating redistricting maps.
However, new methodological developments are needed to properly incorporate the DAS noise generation mechanism into redistricting simulation analyses.
In addition, it remains to be seen whether or not the addition of noise significantly reduces the statistical power to detect racial and partisan gerrymandering.

When considering the fundamental trade-off between privacy protection and data accuracy, it is critical to understand what individual data are at risk.
The decennial Census collects information on individual age, sex, race, relationship to the head of household, and basic housing information, but not other, more sensitive information, such as citizenship, income, and disability status.
The basic demographic variables in the decennial Census play an essential role in public policy, including redistricting---the subject of this paper---and the disbursement of federal and state funds.
Individuals' race/ethnicity is perhaps the most sensitive variable to be protected in the decennial Census microdata.

Indeed, the ability to reveal the race of 17\% of respondents through a microdata reconstruction experiment (using a compendium of five commercial databases) provided key motivation for the Bureau's decision to adopt differential privacy \cite{abowd-reconstruct}.
Combining the Census data with a publicly available voter file, we find that the prediction of individual race is as accurate with the DAS data as with the original Census data.
Although accurate individual-level prediction does not necessarily constitute a violation of \emph{differential} privacy, we believe that this finding needs to be considered when weighing the benefits and costs of privacy protection in the decennial Census.
Our empirical findings on racial imputation accuracy point to the fact that differential privacy does not necessarily prevent accurate prediction of individuals' sensitive information.

As shown in this paper, privacy protection is not free.
It comes with the societal cost of decreasing the accuracy of the decennial Census, which serves as the basis of information in making and evaluating public policy.
Therefore, we must ask what private information we wish to protect, and what cost we are willing to pay for it.
Most importantly, the burden of privacy should not be borne disproportionately by people of certain races or political preferences.

\hypertarget{postscript}{%
\section{Postscript}\label{postscript}}

On August 12, 2021, about six weeks before the planned release, the Census Bureau released the finalized demonstration data for the 2020 Census. The Bureau announced several important changes to the DAS. These changes were based on the comments and feedback submitted during a public comment period in May 2021, where the initial version of this paper was also submitted. First, the Bureau announced a greater privacy loss budget (\(\epsilon = 19.61\)) than for either of the previous releases (\(\epsilon = 12.2\) and \(4.5\)).\footnote{According to the Bureau, they have resolved the problems due to ``geographic bias'' (``the accuracy of population counts being different at larger and smaller geographies'') and ``characteristic bias'' (``counts of racially or ethnically diverse geographies being different than more racially or ethnically homogeneous areas''). \url{https://content.govdelivery.com/accounts/USCENSUS/bulletins/2e5e8a6}}
This constitutes a 1,000-fold increase (on the probability ratio scale) in the leniency of the privacy guarantee since \(\exp(19.61) / \exp(12.2) \approx 1.6 \times 10^3.\)
Second, the Bureau announced several changes to the post-processing algorithm with the goal of reducing biases of the type that we demonstrated in Section 2.
Third, according to the Bureau, they modified the post-processing algorithm to reduce the total error at high levels of geography above the block group. As a side effect, this will likely increase total error
at the block level.

In this section, we repeat our analyses above using the DAS-19.61 data.
The Census Bureau reports that the DAS-19.61 corrects for racial and partisan biases at on-spine geographical units higher than the block group. Unfortunately, we still observe these biases at the VTD-level because the Bureau did not attempt to minimize VTD-level errors as part of their post-processing. This fact has important implications for redistricting simulation analyses, which are typically based on VTDs. We find that although the differences in population counts between the latest DAS and 2010 census data are an order of magnitude smaller than before at the congressional district level, strict population parities may still not be attainable in some cases. In addition, racial and partisan effects of the DAS-19.61 data on simulation analyses remain qualitatively similar to those of the prior DAS releases. It appears that the evaluation of redistricting maps based on the DAS-19.61 can sometimes yield conclusions different from that based on the census data. Finally, like before, the latest DAS does not degrade the overall prediction accuracy of individual voters' race and ethnicity. However, these predictions are sufficiently different to possibly affect the conclusions of simulation analysis for the voting rights of minority groups.

In sum, the latest release of DAS-protected data improves over previous releases
in many ways, but fails to address all of the problems identified in this paper, particularly those affecting the drawing and simulation of districting plans.
Substantial biases remain at the VTD level even after increasing the total privacy loss budget.
These biases would likely not be resolved by any increase in the privacy
loss budget.
Instead, such biases likely come from the decision to maintain accuracy at
geographies other than VTDs and voting precincts, like census tracts.
The imperfect overlap between these boundaries, combined with the increase in errors at small geographic levels like census blocks, could still affect redistricting analyses.
At the same time, using the latest DAS-protected data, we are still able to accurately predict individual's race, which is the most sensitive information of the decennial census.

\hypertarget{racial-and-partisan-biases}{%
\subsection{Racial and Partisan Biases}\label{racial-and-partisan-biases}}

Figures \ref{fig:race-error-19}, \ref{fig:herf-error-19}, and \ref{fig:partisan-error-19}
replicate Figures \ref{fig:race-error}, \ref{fig:herf-error}, and \ref{fig:partisan-error}
on the DAS-19.61 data. For VTDs, we observe the same pattern of bias as before,
albeit around half the magnitude, and despite the Bureau's assurances that racial
biases had been corrected by changes to the post-processing system.

However, it does appear to be the case that the Bureau has largely eliminated these errors for on-spine geographies. Consequently, for larger geographical areas like congressional districts, which can be decomposed as a large number of tracts plus several additional block groups or blocks, the racial biases only manifest in the latter additions, and are relatively insignificant in magnitude.
In contrast, the previous DAS-12.2 contained racial biases for on-spine geographies as well, which were magnified by aggregation and did not disappear, leading to large population shifts in current Congressional districts.
Table \ref{tab:enacted-cd} shows that deviations from 2010 Census totals among all the enacted congressional districts in the states we studied ranged from --2,153 to 2,164 under the DAS-12.2 data, but under the DAS-19.6 data, the range of deviations is only --216 to 319. This constitutes nearly a nearly tenfold improvement.

\hypertarget{simulation-analysis}{%
\subsection{Simulation Analysis}\label{simulation-analysis}}

\hypertarget{population-parity-in-redistricting-1}{%
\subsubsection{Population Parity in Redistricting}\label{population-parity-in-redistricting-1}}

By increasing the privacy loss budget to \(19.61\), we expect total population
errors at large geographic levels like complete districts to be smaller than in
previous releases.
Indeed, the population of Congressional Districts (as of 2019) according to the DAS-19.61 differ from the actual Census counts by an order of magnitude smaller than that according to the DAS-12.2 (Table \ref{tab:enacted-cd}).
However, leaving VTDs ``off-spine'' means that discrepancies
could still be present.

We repeat the parity analysis from Section 3 with the new DAS-19.61
data. As before, we generated 35,000 Louisiana State Senate maps
(5,000 for each of seven tolerances) under each dataset (for 70,000 total)
and measured the fraction of plans that would be rendered invalid.
Figure \ref{fig:la-parity-post} shows the results from this analysis.
Unlike the previous releases of DAS-4.5 and 12.2, the enacted map would still be
valid at an intended population tolerance of 5\% under DAS-19.61.
These errors are lower than those found in Figure \ref{fig:la-parity} for DAS-12.2 and DAS-4.5,
rendering the majority of plans at generous tolerances like 5\% valid in both cases.

However, plans generated with strict parity goals can still be found invalid at high
rates. This means that plans created with a particular parity goal in mind may
in reality exceed that goal in some cases, with the likelihood of such a mistake
happening increasing as the parity tolerance is lowered.

\hypertarget{partisan-effects-on-redistricting-1}{%
\subsubsection{Partisan Effects on Redistricting}\label{partisan-effects-on-redistricting-1}}

Our reanalysis of racial and partisan biases found that biases persist at the VTD level, but generally disappear when VTDs are aggregated to larger, fixed, geographic areas.
Here, we reanalyze the effect of these smaller-scale VTD biases on simulation analyses of partisan and racial gerrymandering.
Do the small-scale errors continue to cause spurious shifts and incorrect conclusions from redistricting simulations, as in Figure \ref{fig:partisan-boxplots}?
Or does the large size of the simulated legislative districts compared to individual VTDs protect against bias?

Unfortunately, as Figure \ref{fig:partisan-boxplots-post} shows, DAS-19.61 data displays qualitatively similar patterns to the DAS-12.2 and DAS-4.5 data.
While for many simulated districts, there is close agreement between the results for Census 2010 and DAS-protected data, for some districts (see Pennsylvania, ordered district 15; and South Carolina State House, ordered districts 111--115 and 121--124) the DAS-based simulations differ by several percentage points, which can shift the direction of a plan's measured partisan bias.

\hypertarget{racial-effects-on-redistricting-1}{%
\subsubsection{Racial Effects on Redistricting}\label{racial-effects-on-redistricting-1}}

The results for racial gerrymandering are similarly troubling.
Figure \ref{fig:race-boxplots-post} shows DAS-19.61 results following the layout of its counterpart in Figure \ref{fig:race-boxplots}.
The same areas for which the DAS-12.2 and DAS-4.5 simulations diverged from the 2010 Census ground truth prove problematic for the DAS-19.61 as well (South Carolina State House, ordered districts 96--110 and 121-124; Mississippi State Senate, ordered districts 49--52; Pennsylvania, ordered district 16).

How should we reconcile these findings with the fact that biases in the overall population totals at the legislative district level appear to have been rectified by DAS-19.61?
We suspect that the simulation process, which constantly makes calculations from and reassigns districts for individual VTDs, is driven more by local considerations than fixed tabulations are.
Analogously to population parity simulations, the aggregated calculations performed by the simulation algorithms and resulting analyses depend on the noisy data itself; this is crucially different than tabulations of existing geographic areas which have been defined without reference to it.

We next examine how the DAS-19.61 affects which VTDs belong to a majority-minority district (MMD). As done in Section 6, we do this through redistricting simulations. We again simulate 50,000 maps using each dataset at a wide variety of constraint strengths that target the creation of majority-minority districts (MMDs). The results are shown in Figure \ref{fig:ebr-race-post}, and they are substantively similar to the results based on the prior demonstration data presented in Figure \ref{fig:ebr-race}. Specifically, we find that some precincts are always contained in an MMD when maps are drawn using the original Census but never under DAS (or vice-versa). As before, the magnitude of such differences are generally larger for higher values of the constraint than for lower values.

\hypertarget{ecological-inference-and-voting-rights-analysis-2}{%
\subsection{Ecological Inference and Voting Rights Analysis}\label{ecological-inference-and-voting-rights-analysis-2}}

\hypertarget{prediction-of-individual-voters-race-and-ethnicity-1}{%
\subsubsection{Prediction of Individual Voters' Race and Ethnicity}\label{prediction-of-individual-voters-race-and-ethnicity-1}}

We examine whether the DAS-19.61 affects the prediction of individual voters' race and ethnicity. Figure \ref{fig:nc-roc-results-19} presents the AUROC results for different racial groups using this final demonstration dataset. We compare the results against those in Figure \ref{fig:nc-roc-results}. We find that, by and large, our conclusions are unchanged. The DAS-19.61 data allows for the prediction of individual voters' race and ethnicity with almost identical levels of accuracy to the 2010 census data. The empirical performance of the BISG methodology based on the DAS-19.61 has a similar pattern: it typically performs about the same for White and Black voters, slightly better for Hispanic voters, and slightly worse for Asian and Other voters. Though the unexpected finding of the DAS-12.2 analysis --- \emph{superior} predictive performance using the privacy-protected data -- is no longer present here, there is also no significant degradation in prediction quality relative to the 2010 census data.

Table \ref{tab:NamePreds-post} reports misclassification rates of the BISG methodology based on the DAS-19.61 data where we assign each individual to the single ethnic group with the highest posterior probability. We can compare these against the analogous results for the 2010 census, DAS-12.2, and DAS-4.5 data given in Tables \ref{tab:lastNamePreds}, \ref{tab:firstNamePreds}, and \ref{tab:middleNamePreds}. The conclusions are largely similar using these metrics too: classification error for individual voters' race and ethnicity is at the same level using the DAS-19.61 data as it is using the 2010 census data.

\hypertarget{ecological-inference-in-the-voting-rights-analysis-1}{%
\subsubsection{Ecological Inference in the Voting Rights Analysis}\label{ecological-inference-in-the-voting-rights-analysis-1}}

We repeat our analysis of the East Ramapo Central School District to examine the effect of the final DAS-19.61 on local redistricting. Using the same geocoded voterfile, we impute race onto the voter file using the BISG with the geographic priors from the DAS-19.61 data. Figure \ref{fig:er-race-scatter-19} displays the imputed races, aggregated to the block level, which is the basic geographic unit for building districts in this case. Consistent with the DAS-12.2 data, the DAS-19.61 data tend to result in overestimates of white voters, and thus underestimates of Black and Hispanic voters.

We find that similar to the previous demonstration data, the block-level undercounts of minority voters do not disappear at the school board ward level in this case. Under DAS-19.61, we find underestimation of majority minority wards in line with findings under DAS-12.2. As shown in Table \ref{tab:er-pair-19}, among sampled districts, majority minority districts are always underestimated in this local case.

\newpage
\FloatBarrier

\hypertarget{appendix-appendix}{%
\appendix}

\renewcommand{\thetable}{A\arabic{table}}
\renewcommand{\thefigure}{A\arabic{figure}}
\setcounter{table}{0}  
\setcounter{figure}{0}

\hypertarget{prediction-of-individual-voters-race-and-ethnicity-accuracy-tables}{%
\section{Prediction of Individual Voters' Race and Ethnicity: Accuracy Tables}\label{prediction-of-individual-voters-race-and-ethnicity-accuracy-tables}}

Here, we give the full set of results for ethnicity classification on the North Carolina dataset.
We assign each individual to the ethnic group with the highest posterior probability, using the Bayesian method described in the main text. Full results are given in Tables \ref{tab:lastNamePreds}, \ref{tab:firstNamePreds}, and \ref{tab:middleNamePreds}.

\begin{table}[ht]
\centering
\begin{tabular}{llrrr}
\toprule
\textbf{Ethnicity} & \textbf{Data}  & \multicolumn{1}{l}{\textbf{Census 2010}} & \multicolumn{1}{l}{\textbf{DAS-4.5}} & \multicolumn{1}{l}{\textbf{DAS-12.2}} \\ \midrule
Overall Error Rate &                & 15.1\%    & 15.6\%  & 15.1\%  \\
White              & False negative & 7.5\%     & 7.9\%   & 7.6\%   \\
White              & False positive & 11.4\%    & 11.7\%  & 11.3\%  \\
Black              & False negative & 29.1\%    & 29.7\%  & 29.1\%  \\
Black              & False positive & 23.6\%    & 24.2\%  & 23.4\%  \\
Hispanic           & False negative & 30.3\%    & 30.7\%  & 27.7\%  \\
Hispanic           & False positive & 29.1\%    & 29.9\%  & 29.0\%  \\
Asian              & False negative & 36.2\%    & 38.2\%  & 34.9\%  \\
Asian              & False positive & 35.4\%    & 35.1\%  & 34.3\%  \\
Other              & False negative & 75.1\%    & 76.0\%  & 75.0\%  \\
Other              & False positive & 35.1\%    & 50.1\%  & 43.9\%  \\ \bottomrule
\end{tabular}
\bigskip
\caption{Overall classification error rate as well as false negative (Type I error) and false positive (Type II error) rates
for White, Black, Latino, Asian, and Other voters using prediction based on geography and \textbf{last names only}. We classify each registered voter to the racial
category with the greatest predicted probability. The columns refer to the different datasets we use to set geographic ethnicity priors at the Census Block level.}
\label{tab:lastNamePreds}
\end{table}

\begin{table}[ht]
\centering
\begin{tabular}{llrrr}
\toprule
\textbf{Ethnicity} & \textbf{Data}  & \multicolumn{1}{l}{\textbf{Census 2010}} & \multicolumn{1}{l}{\textbf{DAS-4.5}} & \multicolumn{1}{l}{\textbf{DAS-12.2}} \\ \midrule
Overall Error Rate &                & 12.1\%      & 12.5\%  & 12.0\%   \\
White              & False negative & 5.4\%       & 5.8\%   & 5.5\%    \\
White              & False positive & 9.3\%       & 9.4\%   & 9.1\%    \\
Black              & False negative & 23.4\%      & 23.9\%  & 23.3\%   \\
Black              & False positive & 17.3\%      & 17.8\%  & 17.1\%   \\
Hispanic           & False negative & 28.2\%      & 28.1\%  & 25.3\%   \\
Hispanic           & False positive & 25.1\%      & 26.1\%  & 24.9\%   \\
Asian              & False negative & 29.2\%      & 31.2\%  & 28.1\%   \\
Asian              & False positive & 30.6\%      & 30.5\%  & 29.5\%   \\
Other              & False negative & 69.8\%      & 71.1\%  & 69.5\%   \\
Other              & False positive & 36.6\%      & 48.0\%  & 43.5\%   \\ \bottomrule
\end{tabular}
\bigskip
\caption{Error rates using prediction based on geography as well as \textbf{last and first names}, for each geographic prior.}
\label{tab:firstNamePreds}
\end{table}

\begin{table}[ht]
\centering
\begin{tabular}{llrrr}
\toprule
\textbf{Ethnicity} & \textbf{Data}  & \multicolumn{1}{l}{\textbf{Census 2010}} & \multicolumn{1}{l}{\textbf{DAS-4.5}} & \multicolumn{1}{l}{\textbf{DAS-12.2}} \\ \midrule
Overall Error Rate &                & 10.0\%      & 10.3\%  & 9.9\%    \\
White              & False negative & 3.9\%       & 4.2\%   & 4.0\%    \\
White              & False positive & 8.0\%       & 8.1\%   & 7.8\%    \\
Black              & False negative & 20.4\%      & 20.9\%  & 20.2\%   \\
Black              & False positive & 12.6\%      & 12.9\%  & 12.3\%   \\
Hispanic           & False negative & 25.2\%      & 24.9\%  & 22.1\%   \\
Hispanic           & False positive & 21.5\%      & 22.6\%  & 21.5\%   \\
Asian              & False negative & 22.8\%      & 24.7\%  & 21.8\%   \\
Asian              & False positive & 26.3\%      & 26.4\%  & 25.3\%   \\
Other              & False negative & 64.2\%      & 66.0\%  & 63.9\%   \\
Other              & False positive & 35.6\%      & 45.1\%  & 41.3\%  \\ \bottomrule
\end{tabular}
\bigskip
\caption{Error rates using prediction based on geography as well as \textbf{last, first, and middle names}, for each geographic prior.}
\label{tab:middleNamePreds}
\end{table}

\FloatBarrier
\clearpage

\renewcommand{\thetable}{B\arabic{table}}
\renewcommand{\thefigure}{B\arabic{figure}}
\setcounter{table}{0}  
\setcounter{figure}{0}

\hypertarget{undercounting-bias-with-das-4.5-data}{%
\section{Undercounting Bias with DAS-4.5 Data}\label{undercounting-bias-with-das-4.5-data}}

\begin{figure}[h]

{\centering \includegraphics[width=0.84\linewidth]{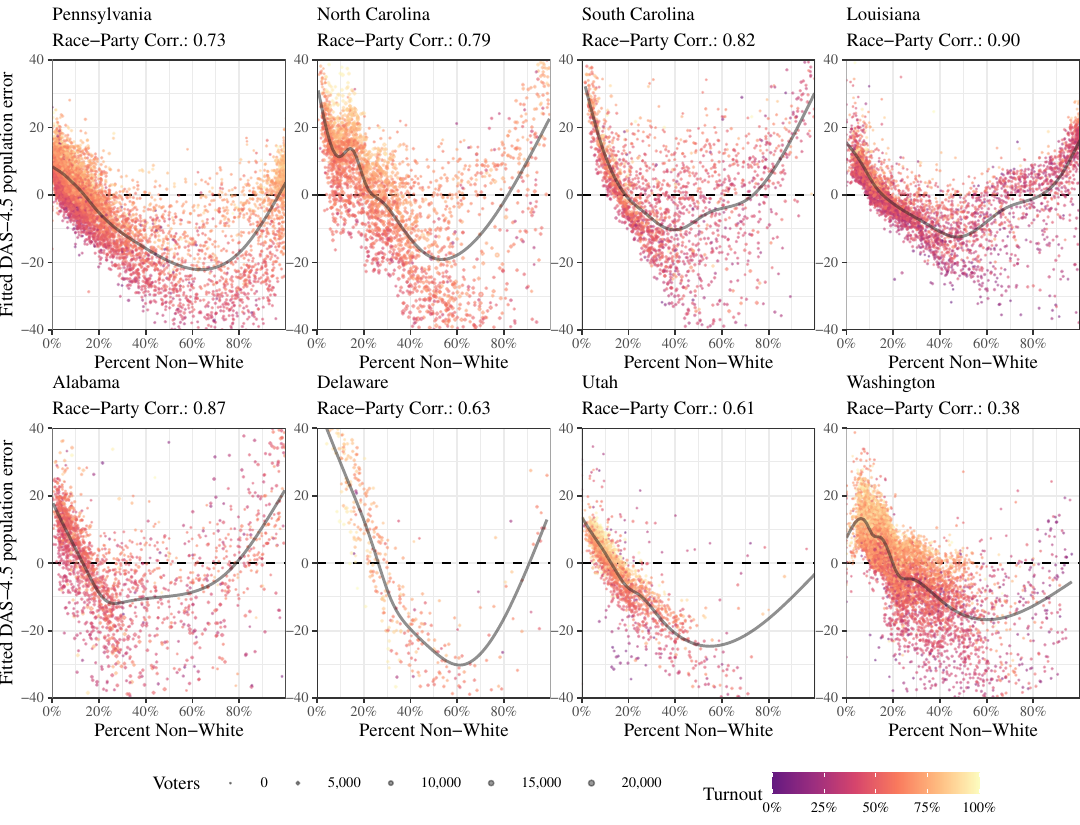} 

}

\caption{DAS-4.5 version of Figure \ref{fig:race-error}.}\label{fig:a-race-error}
\end{figure}

\begin{figure}[h]

{\centering \includegraphics[width=0.84\linewidth]{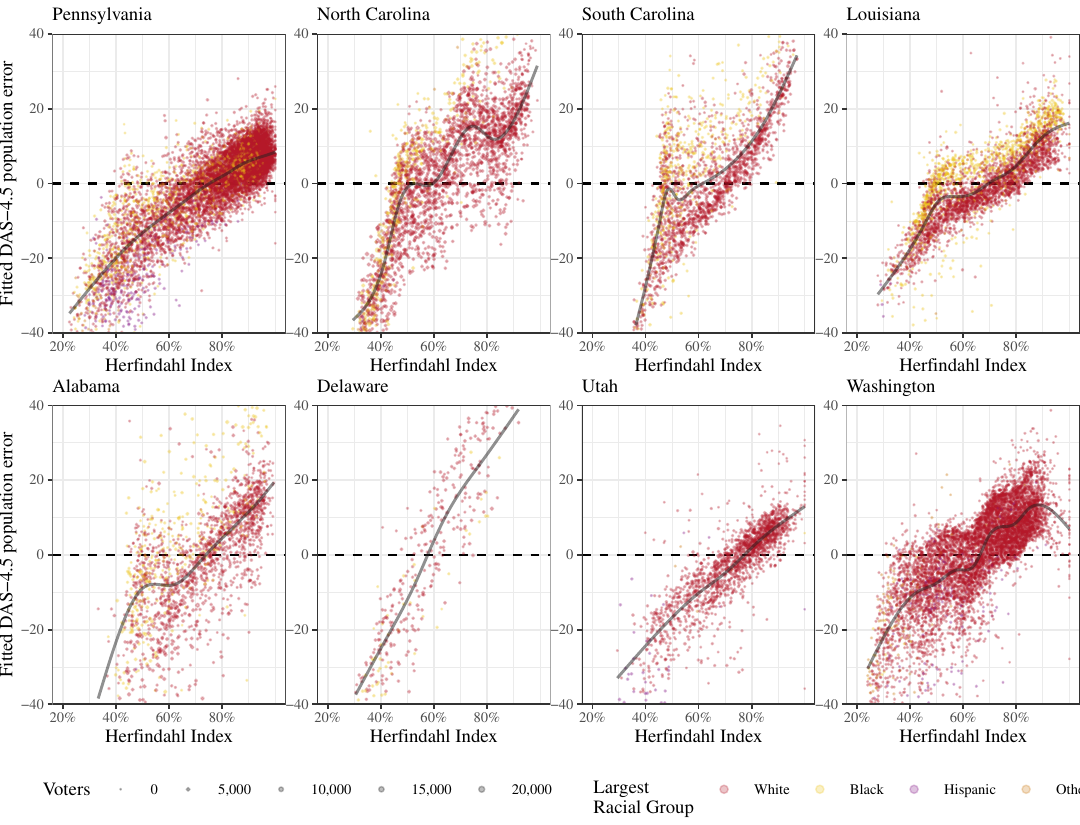} 

}

\caption{DAS-4.5 version of Figure \ref{fig:herf-error}.}\label{fig:a-herf-error}
\end{figure}

\begin{figure}

{\centering \includegraphics[width=0.84\linewidth]{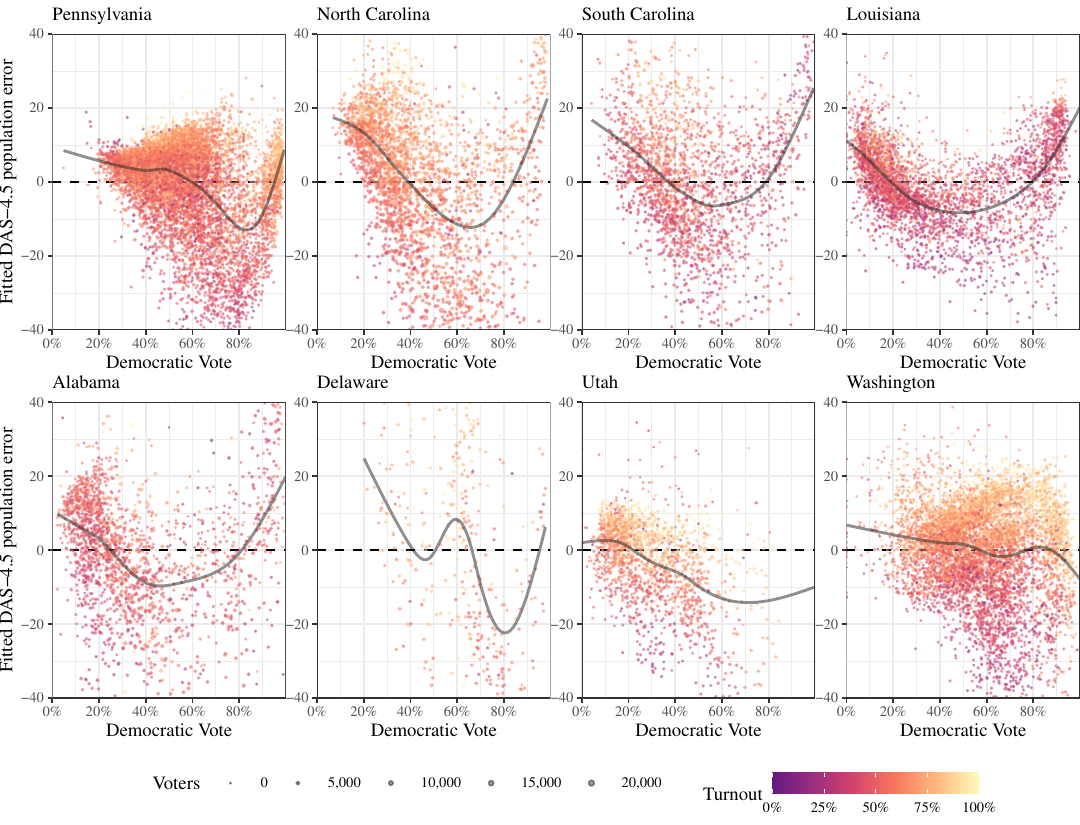} 

}

\caption{DAS-4.5 version of Figure \ref{fig:partisan-error}.}\label{fig:a-partisan-error}
\end{figure}

We replicate three figures from the main paper using the DAS-4.5 data in place of the DAS-12.2 data.
This comparison is helpful for demonstrating that the bias occurs across levels of the privacy budget.
Figure \ref{fig:a-herf-error} follows Figure \ref{fig:herf-error}, Figure \ref{fig:a-partisan-error} follows Figure \ref{fig:partisan-error}, and Figure \ref{fig:a-race-error} follows Figure \ref{fig:race-error}.

\FloatBarrier
\newpage

\renewcommand{\thetable}{C\arabic{table}}
\renewcommand{\thefigure}{C\arabic{figure}}
\setcounter{table}{0}  
\setcounter{figure}{0}

\hypertarget{empirical-results-with-the-das-19.61}{%
\section{Empirical Results with the DAS-19.61}\label{empirical-results-with-the-das-19.61}}

Here we provide re-analyses of the main figures in the text with the DAS-19.61 data released August 12, 2021. The correspondence is as follows:

\begin{table}[h]
\centering
\begin{tabular}{lccc}
\toprule
& DAS-4.5 & DAS-12.2& DAS-19.61\\\midrule
Errors by Race & Fig. \ref{fig:a-race-error} & Fig. \ref{fig:race-error} & Fig. \ref{fig:race-error-19} \\
Errors by Heterogeneity & Fig. \ref{fig:a-herf-error} & Fig. \ref{fig:herf-error} & Fig. \ref{fig:herf-error-19} \\
Errors by Democratic vote & Fig. \ref{fig:a-partisan-error} & Fig. \ref{fig:partisan-error} & Fig. \ref{fig:partisan-error-19} \\
Population Parity (Pennsylvania) & \multicolumn{2}{c}{Fig. \ref{fig:pa-parity}} & - \\
Population Parity (Louisiana) &  \multicolumn{2}{c}{Fig. \ref{fig:la-parity}} & Fig. \ref{fig:la-parity-post}\\
Simulations for Democratic vote &  \multicolumn{2}{c}{Fig. \ref{fig:partisan-boxplots}} & Fig. \ref{fig:partisan-boxplots-post}\\
Simulations for Minority share &  \multicolumn{2}{c}{Fig. \ref{fig:race-boxplots}} & Fig. \ref{fig:race-boxplots-post}\\
Simulations of MMD probability &  \multicolumn{2}{c}{Fig. \ref{fig:ebr-race}} & Fig. \ref{fig:ebr-race-post}\\
Accuracy of BISG &  \multicolumn{2}{c}{Fig. \ref{fig:nc-roc-results}} & Fig. \ref{fig:nc-roc-results-19}\\
False Negatives / False Positives of BISG &  \multicolumn{2}{c}{Tables \ref{tab:lastNamePreds}, \ref{tab:firstNamePreds}, \ref{tab:middleNamePreds}} & Table \ref{tab:NamePreds-post}\\
Block-level Predictions of Race & - & Fig. \ref{fig:er-race-scatter} & Fig. \ref{fig:er-race-scatter-19}\\
Simulations of MMDs in East Ramapo & - & Table \ref{tab:er-pair} & Table \ref{tab:er-pair-19}\\
Comparison with Enacted Maps & Table \ref{tab:enacted-cd} & Table \ref{tab:enacted-cd} & Table \ref{tab:enacted-cd} \\\bottomrule
\end{tabular}
\end{table}

\begin{figure}

{\centering \includegraphics[width=0.84\linewidth]{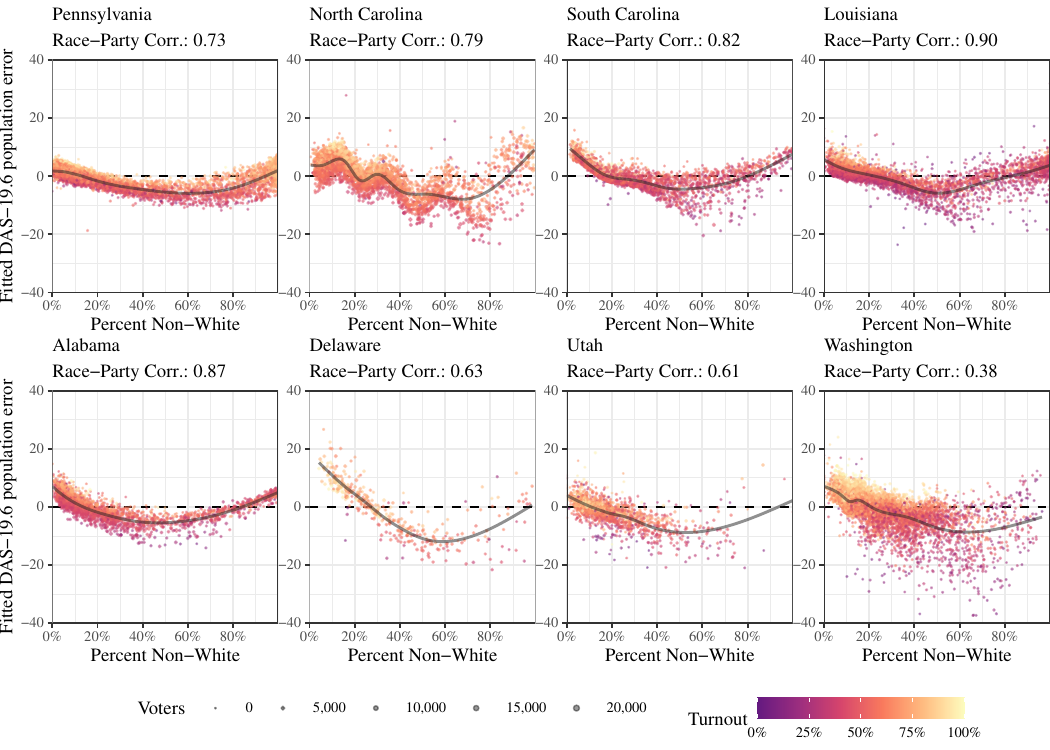} 

}

\caption{DAS-19.61 version of Figure \ref{fig:race-error}.}\label{fig:race-error-19}
\end{figure}

\begin{figure}

{\centering \includegraphics[width=0.84\linewidth]{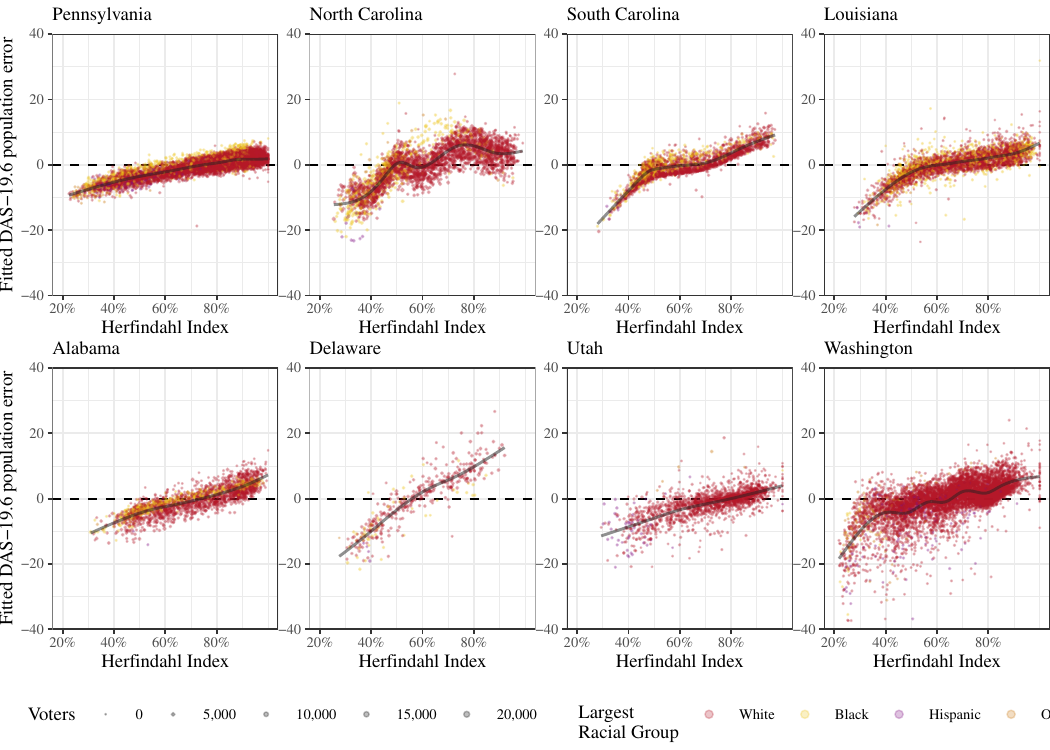} 

}

\caption{DAS-19.61 version of Figure \ref{fig:herf-error}.}\label{fig:herf-error-19}
\end{figure}

\begin{figure}

{\centering \includegraphics[width=0.84\linewidth]{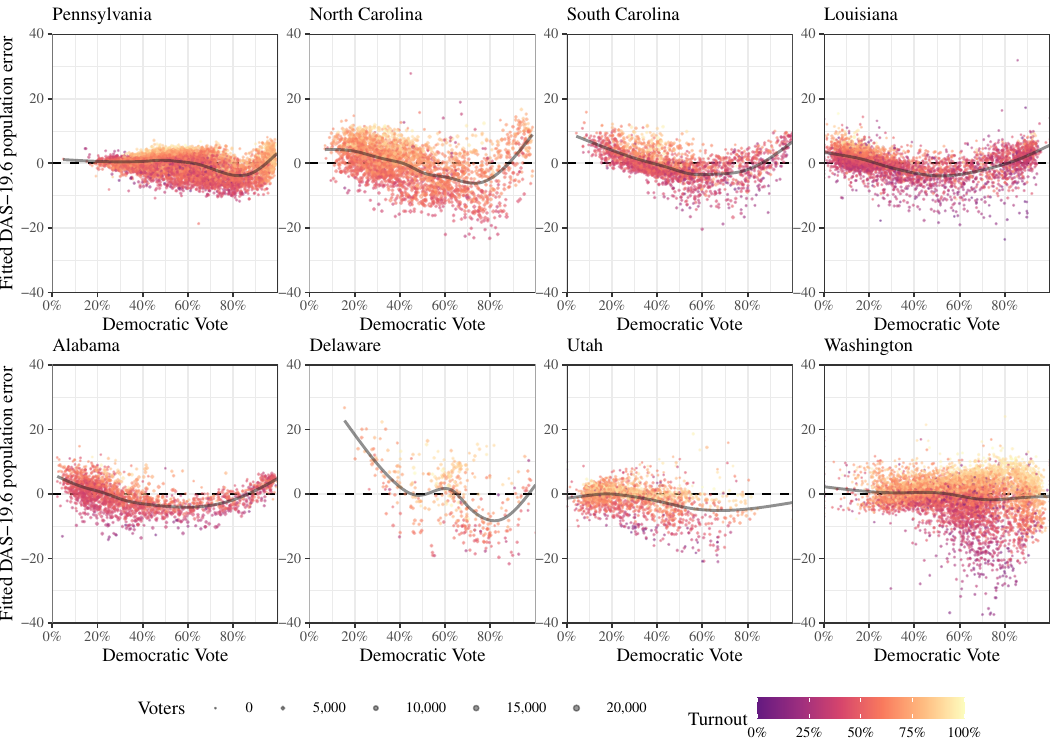} 

}

\caption{DAS-19.61 version of Figure \ref{fig:partisan-error}.}\label{fig:partisan-error-19}
\end{figure}

\begin{figure}[h]

{\centering \includegraphics[width=1\linewidth]{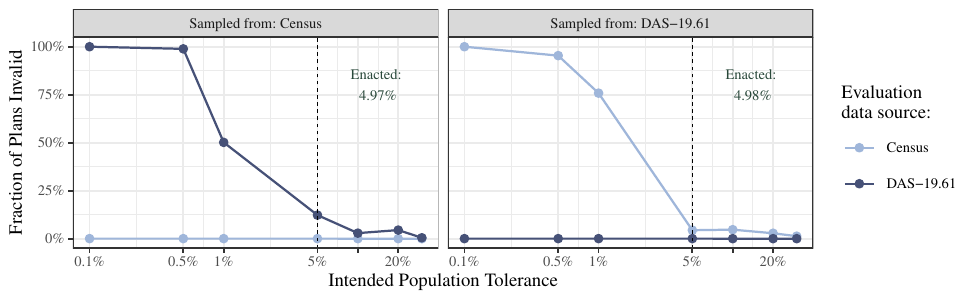} 

}

\caption{DAS-19.61 version of Figure \ref{fig:la-parity}. Fraction of Louisiana State Senate plans simulated under one data source with a population parity constraint which are invalid when measured under another. The horizontal axis shows the tolerance constraint for the original simulation on the log10 scale. The vertical axis shows the percent of plans that exceed the intended tolerance according to the evaluation data.  The dashed line shows the maximum deviation from parity of the enacted 2010 map.}\label{fig:la-parity-post}
\end{figure}

\begin{figure}

{\centering \includegraphics[width=1\linewidth]{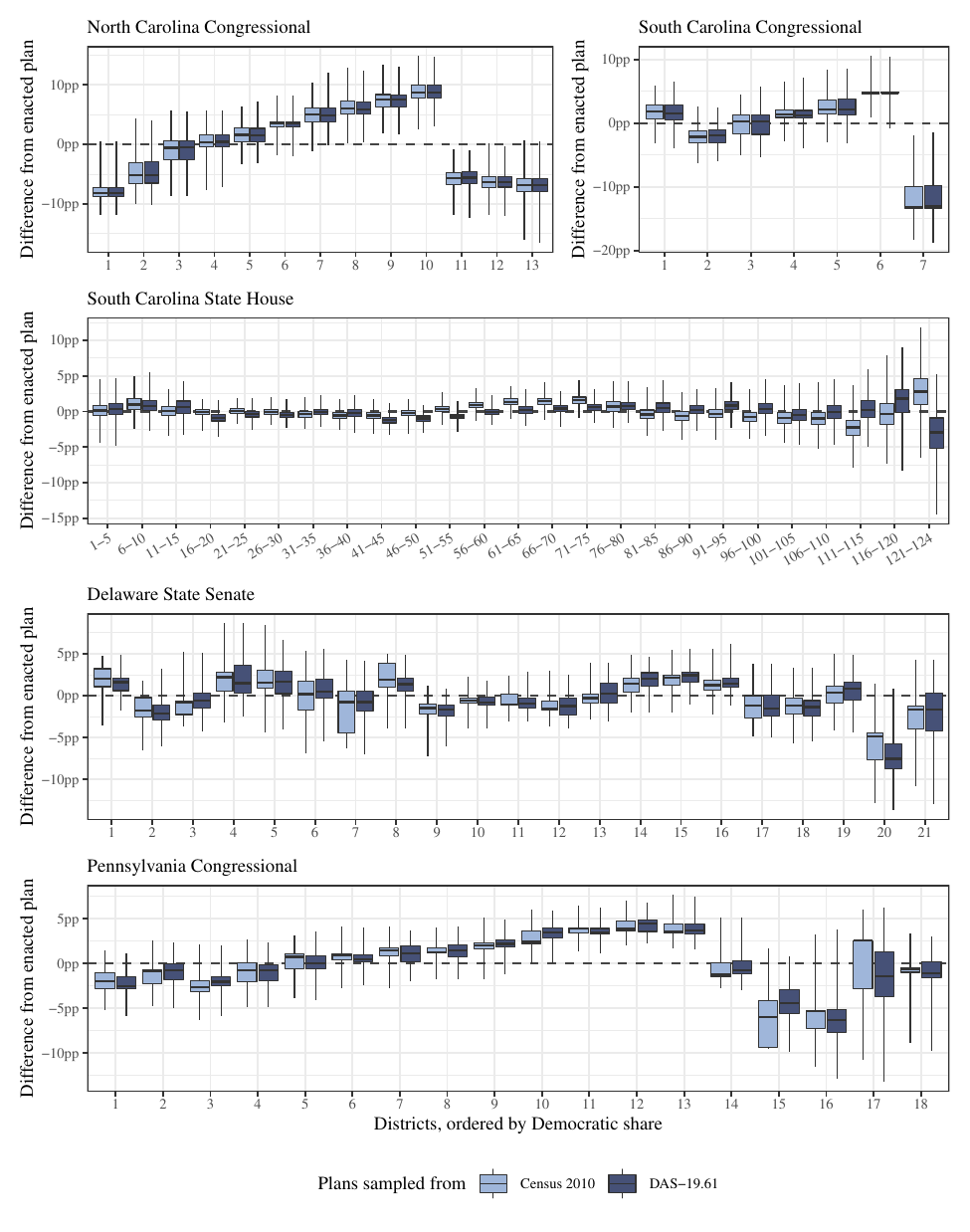} 

}

\caption{DAS-19.61 version of Figure \ref{fig:partisan-boxplots}.}\label{fig:partisan-boxplots-post}
\end{figure}

\begin{figure}

{\centering \includegraphics[width=1\linewidth]{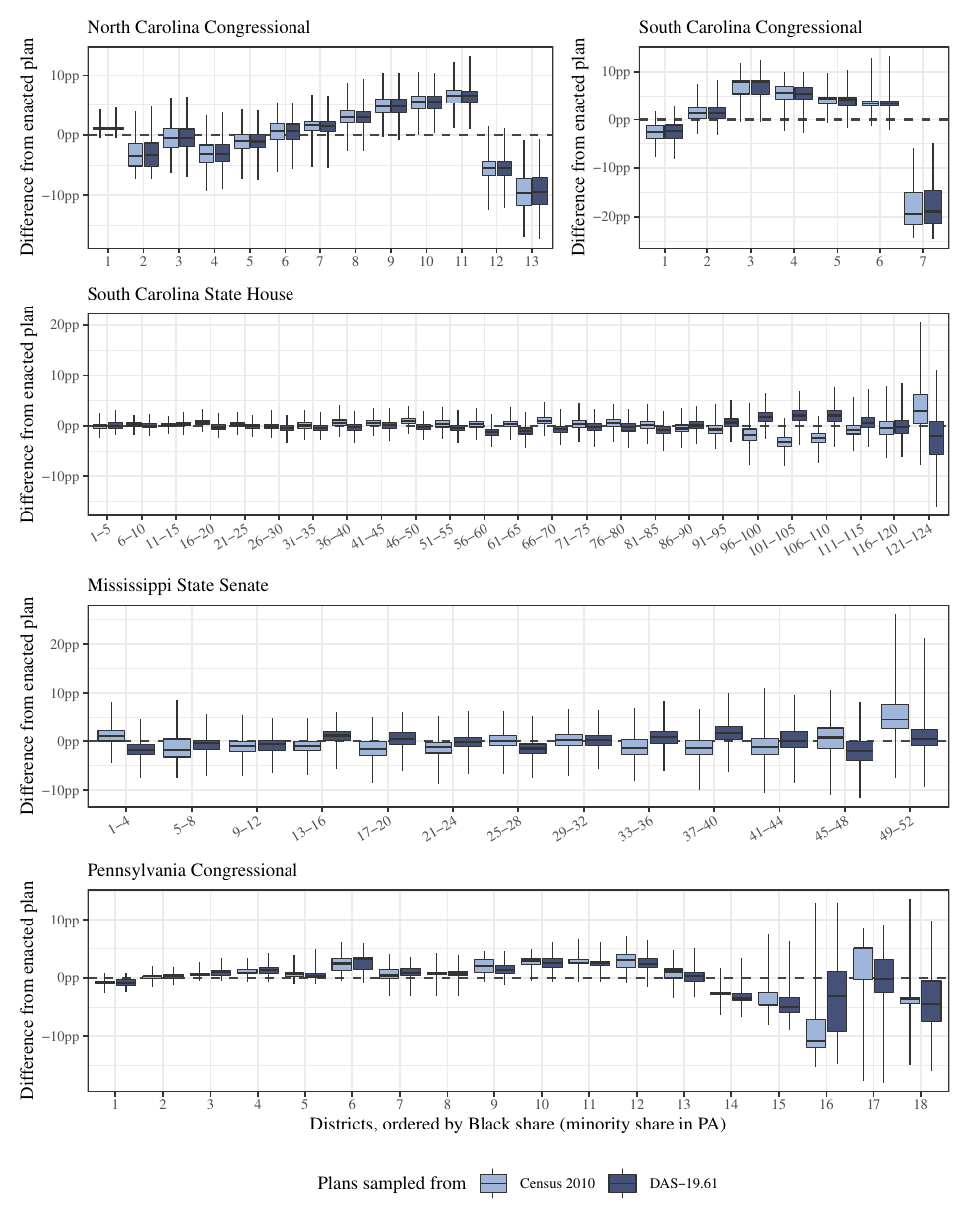} 

}

\caption{DAS-19.61 version of Figure \ref{fig:race-boxplots}.}\label{fig:race-boxplots-post}
\end{figure}

\begin{figure}

{\centering \includegraphics[width=1\linewidth]{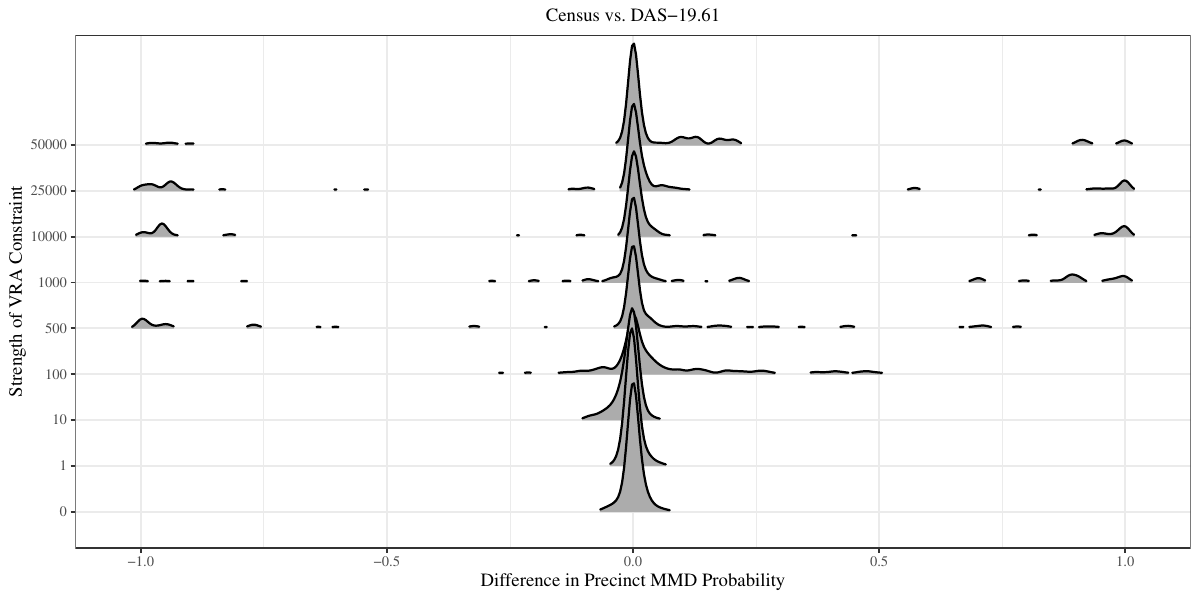} 

}

\caption{DAS-19.61 version of Figure \ref{fig:ebr-race}. The calculated probability of being assigned to a majority-minority district can be much higher or lower for individual precincts, and these differences grow as a constraint encouraging the formation of MMDs is strengthened.}\label{fig:ebr-race-post}
\end{figure}

\FloatBarrier
\newpage

\begin{figure}

{\centering \includegraphics[width=5.5in]{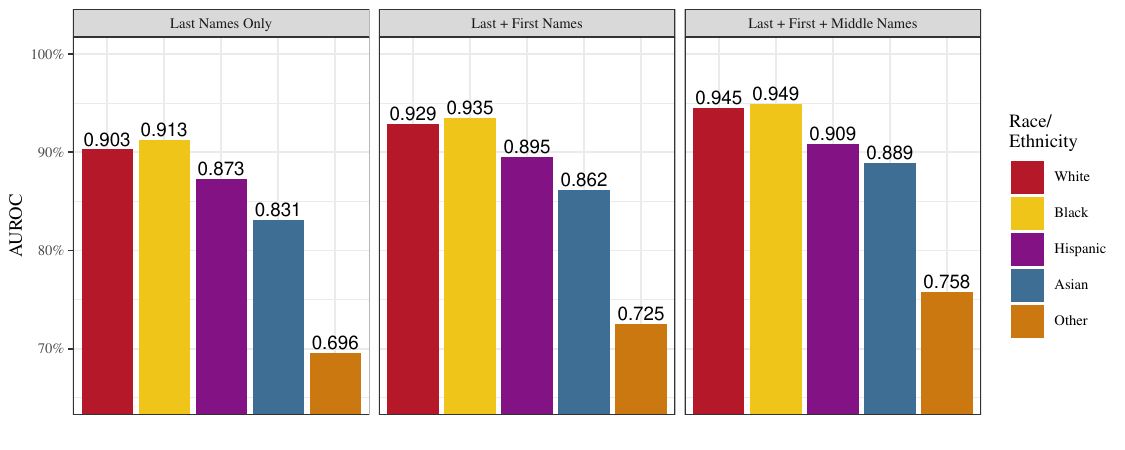} 

}

\caption{DAS-19.61 version of Figure \ref{fig:nc-roc-results}. Area under the Receiver Operating Characteristic Curve (AUROC) percentage values for the prediction of each individual voter's race and ethnicity using the North Carolina voter file, with geographic priors given by the DAS-19.61 dataset.}\label{fig:nc-roc-results-19}
\end{figure}

\FloatBarrier

\begin{table}[ht]
\centering
\begin{tabular}{llrrr}
\toprule
\textbf{Ethnicity} & \textbf{Data}  & \multicolumn{1}{l}{\textbf{Last}} & \multicolumn{1}{l}{\textbf{Last, First}} & \multicolumn{1}{l}{\textbf{Last, First, Middle}}  \\ \midrule
Overall Error Rate &                & 15.5\%      & 12.4\%  & 10.2\%   \\
White              & False negative & 7.9\%       & 5.8\%   &  4.2\%  \\
White              & False positive & 11.5\%       & 9.3\%  & 7.9\%  \\
Black              & False negative & 29.5\%      & 23.7\% & 20.7\%  \\
Black              & False positive & 24.2\%       & 17.7\% & 12.8\%  \\
Hispanic           & False negative & 30.2\%      & 27.4\% & 24.1\%  \\
Hispanic           & False positive & 29.1\%      & 25.1\% & 21.8\%  \\
Asian              & False negative & 37.1\%      & 30.2\% & 23.8\%  \\
Asian              & False positive & 34.5\%      & 29.7\% & 25.7\%  \\
Other              & False negative & 74.7\%      & 69.5\%  & 64.0\%  \\
Other              & False positive & 48.8\%      & 47.6\% & 45.0\% \\ \bottomrule
\end{tabular}
\bigskip
\caption{DAS-19.61 version of Tables \ref{tab:lastNamePreds}, \ref{tab:firstNamePreds}, \ref{tab:middleNamePreds}. Overall classification error rate as well as false negative (Type I error) and false positive (Type II error) rates
for White, Black, Latino, Asian, and Other voters using prediction based on geographic priors derived from the DAS-19.61 data.}
\label{tab:NamePreds-post}
\end{table}

\begin{figure}

{\centering \includegraphics[width=0.9\linewidth]{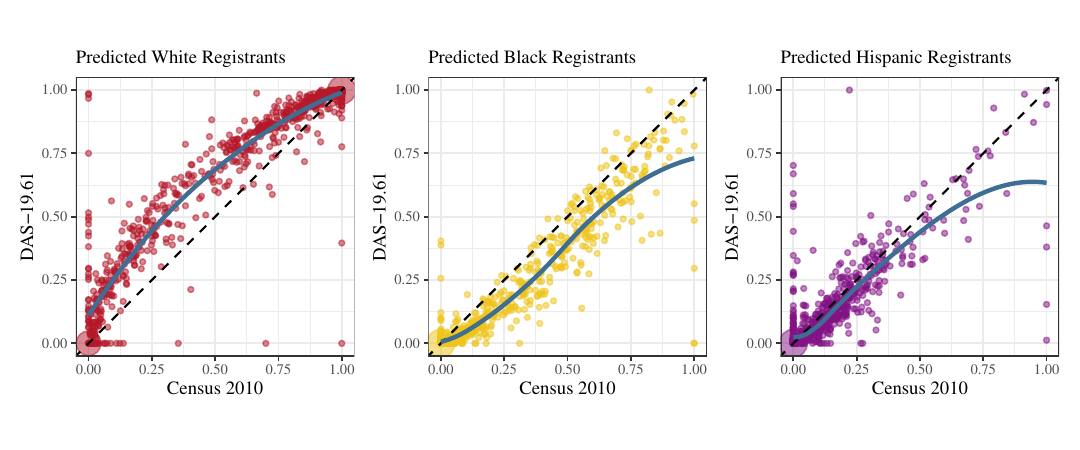} 

}

\caption{DAS-10.61 version of Figure \ref{fig:er-race-scatter}. Imputed Racial Registrants by Census Blocks. The x-axis represents the percent of a group, as measured by the most likely race from racial imputation using the Census 2010 data. The y-axis represents the corresponding imputation using the DAS-19.61 data.}\label{fig:er-race-scatter-19}
\end{figure}

\begin{table}

\caption{\label{tab:er-pair-19}DAS-19.61 version of Table \ref{tab:er-pair}. East Ramapo MMDs under Census 2010 and DAS-19.61 data. The noise introduced in the DAS-19.61 leads us to undercount the number of majority minority districts in many plans, but never to overcount them.}
\centering
\begin{tabular}[t]{c>{\centering\arraybackslash}p{1cm}>{\centering\arraybackslash}p{1cm}>{\centering\arraybackslash}p{1cm}>{\centering\arraybackslash}p{1cm}r}
\toprule
\multicolumn{1}{c}{ } & \multicolumn{4}{c}{Number of MMDs from DAS-19.61} & \multicolumn{1}{c}{ } \\
\cmidrule(l{3pt}r{3pt}){2-5}
Census 2010 & 0 & 1 & 2 & 3 & Plans\\
\midrule
0 & \textbf{100\%} & 0 & 0 & 0 & 4\\
1 & 26 & \textbf{74} & 0 & 0 & 3,435\\
2 & 66 & 14 & \textbf{20} & 0 & 6,462\\
3 & 0 & 42 & 58 & \textbf{0} & 99\\
\bottomrule
\multicolumn{6}{l}{\textsuperscript{} Note: Percentages add to 100\% by row.}\\
\end{tabular}
\end{table}

\begin{table}[p]
\caption{Difference in the population of existing Congressional Districts as computed by three DAS populations, relative to the enacted Congressional District Map. For each table and each state we show the minimum, median, maximum, and standard deviation of the difference in population. All numbers are rounded to whole numbers. \label{tab:enacted-cd}}

\centering

\bigskip

\textbf{DAS-4.5}

\begin{tabular}[t]{lrrrrrr}
\toprule
State & Min & Median & Mean & Max & SD & CDs\\
\midrule
Alabama & -597 & -2 & 0 & 865 & 444 & 7\\
Delaware & 0 & 0 & 0 & 0 &  & 1\\
Louisiana & -1,221 & 35 & 0 & 669 & 660 & 6\\
Mississippi & -262 & 12 & 0 & 237 & 210 & 4\\
North Carolina & -1,301 & -9 & 0 & 1,659 & 657 & 13\\
Pennsylvania & -3,996 & 0 & 0 & 3,805 & 1,453 & 18\\
South Carolina & -467 & -173 & 0 & 1,318 & 633 & 7\\
Utah & -1,303 & 166 & 0 & 971 & 1,065 & 4\\
Washington & -3,128 & 54 & 0 & 2,784 & 1,617 & 10\\
\bottomrule
\end{tabular}

\bigskip

\textbf{DAS-12.2}

\begin{tabular}[t]{lrrrrrr}
\toprule
State & Min & Median & Mean & Max & SD & CDs\\
\midrule
Alabama & -405 & -2 & 0 & 483 & 270 & 7\\
Delaware & 0 & 0 & 0 & 0 &  & 1\\
Louisiana & -471 & 30 & 0 & 314 & 259 & 6\\
Mississippi & -210 & 27 & 0 & 156 & 159 & 4\\
North Carolina & -601 & -18 & 0 & 680 & 277 & 13\\
Pennsylvania & -2,153 & 1 & 0 & 2,164 & 765 & 18\\
South Carolina & -287 & -6 & 0 & 485 & 256 & 7\\
Utah & -328 & -18 & 0 & 363 & 291 & 4\\
Washington & -2,099 & 98 & 0 & 1,413 & 1,075 & 10\\
\bottomrule
\end{tabular}

\bigskip

\textbf{DAS-19.61}

\begin{tabular}[t]{lrrrrrr}
\toprule
State & Min & Median & Mean & Max & SD & CDs\\
\midrule
Alabama & -163 & 17 & 0 & 161 & 102 & 7\\
Delaware & 0 & 0 & 0 & 0 &  & 1\\
Louisiana & -196 & 56 & 0 & 96 & 115 & 6\\
Mississippi & -150 & 36 & 0 & 79 & 104 & 4\\
North Carolina & -216 & 10 & 0 & 319 & 133 & 13\\
Pennsylvania & -122 & -12 & 0 & 121 & 61 & 18\\
South Carolina & -100 & 8 & 0 & 89 & 72 & 7\\
Utah & -42 & 6 & 0 & 29 & 30 & 4\\
Washington & -143 & 13 & 0 & 200 & 92 & 10\\
\bottomrule
\end{tabular}
\end{table}

\FloatBarrier
\clearpage

\bibliographystyle{unsrt}
\bibliography{references.bib}

\begin{thebibliography}{10}

\bibitem{abow:etal:20}
John~M. Abowd, Gary~L. Benedetto, Simson~L. Garfinkel, Scot~A. Dahl, Aref~N.
  Dajani, Matthew Graham, Michael~B. Hawes, Vishesh Karwa, Daniel Kifer, Hang
  Kim, Philip Leclerc, Aashwin Machanavajjhala, Jerome~P. Reiter, Rolando
  Rodriguez, Ian~M. Schmutte, William~N. Sexton, Phyllis~E. Singer, and Lars
  Vilhuber.
\newblock The modernization of statistical disclosure limitation at the {U.S.}
  {C}ensus {B}ureau.
\newblock 2020.

\bibitem{rugg:etal:19}
Steven Ruggles, Catherine Fitch, Diana Magnuson, and Jonathan Schroeder.
\newblock Differential privacy and census data: Implications for social and
  economic research.
\newblock {\em AEA Papers and Proceedings}, 109:403--408, 2019.

\bibitem{nas:20}
{National Academies of Sciences, Engineering, and Medicine}.
\newblock {2020 Census Data Products: Data Needs and Privacy Considerations},
  2020.

\bibitem{ripe:kugl:rugg:20}
David Van~Riper, Tracy Kugler, and Steven Ruggles.
\newblock Disclosure avoidance in the {C}ensus {B}ureau’s 2010 demonstration
  data product.
\newblock In {\em Privacy in Statistical Databases}, Lecture Notes in Computer
  Science, pages 353--368. Springer International Publishing, 2020.

\bibitem{sant:etal:20}
Alexis~R. Santos-Lozada, Jeffrey~T. Howard, and Ashton~M. Verdery.
\newblock {How differential privacy will affect our understanding of health
  disparities in the United States}.
\newblock {\em Proceedings of the National Academy of Sciences},
  117(24):13405--13412, 2020.

\bibitem{muel:sant:21}
Tom Mueller and Alexis~R. Santos-Lozada.
\newblock Proposed {U.S.} census bureau differential privacy method is biased
  against rural and non-white populations, Jun 2021.

\bibitem{chen2013}
Jowei Chen and Jonathan Rodden.
\newblock Unintentional gerrymandering: Political geography and electoral bias
  in legislatures.
\newblock {\em Quarterly Journal of Political Science}, 8(3):239--269, 2013.

\bibitem{cart:etal:19}
Daniel Carter, Gregory Herschlag, Zach Hunter, and Jonathan Mattingly.
\newblock A merge-split proposal for reversible {M}onte {C}arlo {M}arkov chain
  sampling of redistricting plans.
\newblock {\em arXiv preprint arXiv:1911.01503}, 2019.

\bibitem{deford2019}
Daryl DeFord, Moon Duchin, and Justin Solomon.
\newblock Recombination: A family of {M}arkov chains for redistricting.
\newblock {\em Harvard Data Science Review}, 2021.

\bibitem{autry2020}
Eric Autry, Daniel Carter, Gregory Herschlag, Zach Hunter, and Jonathan~C.
  Mattingly.
\newblock Multi-scale merge-split {M}arkov chain {M}onte {C}arlo for
  redistricting.
\newblock {\em arXiv preprint arXiv:2008.08054}, 2020.

\bibitem{fifi:etal:20}
Benjamin Fifield, Michael Higgins, Kosuke Imai, and Alexander Tarr.
\newblock Automated redistricting simulator using {Markov chain Monte Carlo}.
\newblock {\em Journal of Computational and Graphical Statistics},
  29(4):715--728, 2020.

\bibitem{mcca:imai:20}
Cory McCartan and Kosuke Imai.
\newblock Sequential {M}onte {C}arlo for sampling balanced and compact
  redistricting plans.
\newblock {\em arXiv preprint arXiv:2008.06131}, 2020.

\bibitem{kenn:etal:21}
Christopher~T. Kenny, Cory McCartan, Ben Fifield, and Kosuke Imai.
\newblock {redist}: Simulation methods for legislative redistricting.
\newblock Available at The Comprehensive R Archive Network (CRAN), 2021.

\bibitem{fisc:frem:06}
Kevin Fiscella and Allen~M. Fremont.
\newblock Use of geocoding and surname analysis to estimate race and ethnicity.
\newblock {\em Health Services Research}, 41(4p1):1482--1500, August 2006.

\bibitem{elli:etal:09}
Marc~N. Elliott, Peter~A. Morrison, Allen Fremont, Daniel~F. {McCaffrey},
  Philip Pantoja, and Nicole Lurie.
\newblock Using the {C}ensus {B}ureau's surname list to improve estimates of
  race/ethnicity and associated disparities.
\newblock {\em Health Services and Outcomes Research Methodology}, 9(2):69--83,
  2009.

\bibitem{imai:khan:16}
Kosuke Imai and Kabir Khanna.
\newblock Improving ecological inference by predicting individual ethnicity
  from voter registration record.
\newblock {\em Political Analysis}, 24(2):263--272, Spring 2016.

\bibitem{MGGG:21}
Aloni Cohen, Moon Duchin, JN~Matthews, and Bhushan Suwal.
\newblock {Census TopDown: The Impacts of Differential Privacy on
  Redistricting}.
\newblock In Katrina Ligett and Swati Gupta, editors, {\em 2nd Symposium on
  Foundations of Responsible Computing (FORC 2021)}, volume 192 of {\em Leibniz
  International Proceedings in Informatics (LIPIcs)}, pages 5:1--5:22,
  Dagstuhl, Germany, 2021. Schloss Dagstuhl -- Leibniz-Zentrum f{\"u}r
  Informatik.

\bibitem{geomander}
Christopher~T. Kenny.
\newblock {geomander}: Geographic tools for studying gerrymandering.
\newblock Available at The Comprehensive R Archive Network (CRAN), 2021.

\bibitem{ppmf}
Christopher~T. Kenny.
\newblock {ppmf}: ppmf: Read census privacy protected microdata files, 2021.
\newblock R package version 0.0.3.

\bibitem{vest}
VEST.
\newblock {Voting and Election Science Team Dataverse}, 2021.
\newblock Available at
  \url{https://dataverse.harvard.edu/dataverse/electionscience}.

\bibitem{ncsl:12}
National~Conference of~State~Legislatures.
\newblock 2010 redistricting deviation table, 2012.

\bibitem{census-das-targets}
United States~Census Bureau.
\newblock Meeting redistricting data requirements: Accuracy targets, 2021.
\newblock Available at
  \url{https://content.govdelivery.com/accounts/USCENSUS/bulletins/2cb745b}.

\bibitem{good:53}
Leo~A. Goodman.
\newblock Ecological regressions and behavior of individuals.
\newblock {\em American Sociological Review}, 18(6):663--664, 1953.

\bibitem{king:rose:tann:04}
Gary King, Ori Rosen, and Martin Tanner, editors.
\newblock {\em Ecological Inference: New Methodological Strategies}.
\newblock Cambridge University Press, 2004.

\bibitem{rosenbluth1955measures}
Gideon Rosenbluth.
\newblock {Measures of Concentration}.
\newblock In {\em {Business Concentration and Price Policy}}, pages 57--99.
  Princeton University Press, 1955.

\bibitem{chen:rodd:15}
Jowei Chen and Jonathan Rodden.
\newblock Cutting through the thicket: Redistricting simulations and the
  detection of partisan gerrymanders.
\newblock {\em Election Law Journal}, 14(4):331--345, 2015.

\bibitem{hers:etal:20}
Gregory Herschlag, Han~Sung Kang, Justin Luo, Christy~Vaughn Graves, Sachet
  Bangia, Robert Ravier, and Jonathan~C. Mattingly.
\newblock Quantifying gerrymandering in north carolina.
\newblock {\em Statistics and Public Policy}, 7(1):30--38, 2020.

\bibitem{poli-sci-amicus:19}
{Brief of Amici Curiae: Political science professors in support of appellees
  and affirmance}.
\newblock {\em Rucho v. Common Cause}, 2019.

\bibitem{pegd:rodd:wang:19}
{Brief of Amici Curiae: Professors Wesley Pegden, Jonathan Rodden, and Samuel
  S.-H. Wang in support of appellees}.
\newblock {\em Rucho v. Common Cause}, 2019.

\bibitem{issacharoff1997standing}
Samuel Issacharoff and Pamela~S. Karlan.
\newblock Standing and misunderstanding in voting rights law.
\newblock {\em Harvard Law Review}, 111:2276, 1997.

\bibitem{voic:18}
Ioan Voicu.
\newblock Using first name information to improve race and ethnicity
  classification.
\newblock {\em Statistics and Public Policy}, 5(1):1--13, 2018.

\bibitem{censusSurnameList}
{United States Census Bureau}.
\newblock Frequently occurring surnames from the 2010 {C}ensus.
\newblock
  \url{https://www.census.gov/topics/population/genealogy/data/2010_surnames.html},
  2016.

\bibitem{lobo2008auc}
Jorge~M. Lobo, Alberto Jim{\'e}nez-Valverde, and Raimundo Real.
\newblock {AUC}: a misleading measure of the performance of predictive
  distribution models.
\newblock {\em Global Ecology and Biogeography}, 17(2):145--151, 2008.

\bibitem{censusxy}
Christopher Prener and Branson Fox.
\newblock {\em censusxy: Access the U.S. Census Bureau's Geocoding A.P.I.
  System}, 2021.
\newblock R package version 1.0.1.

\bibitem{pren:fox:21}
Christopher Prener and Branson Fox.
\newblock Creating open source composite geocoders: Pitfalls and opportunities.
\newblock {\em Transactions in GIS}, 2021.

\bibitem{wru}
Kabir Khanna and Kosuke Imai.
\newblock {\em wru: Who are You? Bayesian Prediction of Racial Category Using
  Surname and Geolocation}, 2021.
\newblock R package version 0.1-12.

\bibitem{abowd-reconstruct}
John~M. Abowd.
\newblock {Affidavit Declaration of John M. Abowd}.
\newblock Defendants’ response in opposition to combined motion for a
  preliminary injunction and petition for a writ of mandamus, \emph{Alabama v.
  U.S. Dep’t of Commerce}, No. 3: 21-cv-211-RAH-KFP, 2021.
\newblock Available at \url{https://perma.cc/E6TJ-SSTU}.

\end{thebibliography}

\end{document}